\documentclass[]{aa}
\usepackage{color}
\usepackage{graphicx}
\usepackage{longtable}
\usepackage{lscape}
\usepackage{multirow}
\usepackage{natbib}
\usepackage[scaled]{helvet}
\usepackage[varg]{txfonts}
\usepackage{url}
\usepackage{xspace}
\bibpunct{(}{)}{;}{a}{}{,}
\newcounter{Rco}

\newcommand{\Ionst}[1]{\setcounter{Rco}{#1}\Roman{Rco}}
\newcommand{\Ion}[2]{\mbox{#1\,{\scriptsize\Ionst{#2}}}}

\newcommand{\logg}{\mbox{$\log g$}\xspace}

\newcommand{\sA}[1]{\mbox{(Fig.\,\ref{#1})}}

\newcommand{\Teff}{\mbox{$T_\mathrm{eff}$}\xspace}

\newcommand{\vrad}{\mbox{$v_\mathrm{rad}$}\xspace}

\newcommand{\Msol}{$M_\odot$\xspace}
\newcommand{\two}{\multicolumn{2}{c}{}}

\newcommand{\re}{\object{RE\,0503$-$289}\xspace}
\newcommand{\wdnull}{\object{WD\,0111$+$002}\xspace}
\newcommand{\pgsieben}{\object{PG\,1707$+$427}\xspace}
\newcommand{\pgnull}{\object{PG\,0109$+$111}\xspace}
\begin{document}

\title{Search for trans-iron elements in hot, helium-rich white dwarfs \\ with the HST Cosmic Origins Spectrograph\thanks
           {Based on observations with the NASA/ESA Hubble Space Telescope, obtained at the Space Telescope Science 
            Institute, which is operated by the Association of Universities for Research in Astronomy, Inc., under 
            NASA contract NAS5-26666.
           } \thanks{Based on observations made with the NASA-CNES-CSA Far
    Ultraviolet Spectroscopic Explorer.} 
      }
\titlerunning{Search for trans-iron elements in hot, helium-rich white dwarfs with HST/COS}

\author{D\@. Hoyer\inst{1}
        \and
        T\@. Rauch\inst{1}
        \and
        K\@. Werner\inst{1}
        \and
        J.W\@. Kruk\inst{2}
       }

\institute{Institute for Astronomy and Astrophysics,
           Kepler Center for Astro and Particle Physics,
           Eberhard Karls University,
           Sand 1,
           72076 T\"ubingen,
           Germany \\
           \email{werner@astro.uni-tuebingen.de}
\and
           NASA Goddard Space Flight Center, Greenbelt, MD\,20771, USA
}

\date{Received 4 December 2017 / Accepted 4 January 2018}

\abstract {The metal abundances in the atmospheres of hot white dwarfs
  (WDs) entering the cooling sequence are determined by the preceding
  Asymptotic Giant Branch (AGB) evolutionary phase and, subsequently,
  by the onset of gravitational settling and radiative levitation. In
  this paper, we investigate three hot He-rich WDs, which are believed
  to result from a late He-shell flash. During such a flash, the
  He-rich intershell matter is dredged up and dominates the surface
  chemistry. Hence, in contrast to the usual H-rich WDs, their spectra
  allow direct access to s-process element abundances in the
  intershell that were synthesized during the AGB stage. In order to
  look for trans-iron group elements (atomic number $Z>29$), we
  performed a non-local thermodynamic equilibrium model atmosphere
  analysis of new ultraviolet spectra taken with the Cosmic Origins
  Spectrograph aboard the \emph{Hubble} Space Telescope. One of our
  program stars is of PG\,1159 spectral type; this star, \pgsieben,
  has  effective temperature \Teff $= 85\,000$\,K, and surface gravity
  \logg $=7.5$. The two other stars are DO white dwarfs: \wdnull\ has
  \Teff $= 58\,000$\,K and \logg $=7.7$, and \pgnull\ has \Teff $=
  70\,000$\,K and \logg $=8.0$. These stars trace the onset of element
  diffusion during early WD evolution. While zinc is the only
  trans-iron element we could detect in the PG\,1159 star, both DOs
  exhibit lines from Zn, Ga, Ge, Se; one  additionally exhibits lines
  from Sr, Sn, Te, and I and the other from  As. Generally, the
  trans-iron elements are very abundant in the DOs, meaning that
  radiative levitation must be acting. Most extreme is the almost six
  orders of magnitude oversolar abundance of tellurium in \pgnull. In
  terms of mass fraction, it is the most abundant metal in the
  atmosphere. The two DOs join the hitherto unique hot DO \re, in
  which 14 trans-iron elements had even been identified.}


\maketitle

\begin{table*}[t]
\begin{center}
\caption{Observation log of our three program stars.\tablefootmark{a} }
\label{tab:obs} 
\small
\begin{tabular}{ccccccccc}
\hline 
\hline 
\noalign{\smallskip}
Star    & Instrument & Dataset   & Grating  & $R$        &$\lambda$/\AA         & $t_{\rm exp}$/s&Date&PI  \\
\hline 
\noalign{\smallskip}
\wdnull & HST/COS    & LCK201010 & G130M    &13000--19000& 1070--1210, 1224--1365& 8017 & 2015-11-17 &Werner\\
        & HST/GHRS   & Z3GM0304T & G160M    & 18000      & 1228--1264           & 4570 & 1996-11-11 &Werner\\
        &  FUSE      & A0130301000&         & 20000      & 915--1188            & 5515 & 2001-08-07 &Dreizler\\
\noalign{\smallskip}
\pgnull & HST/COS    & LCK202010 & G130M    &13000--19000& 1070--1210, 1224--1365& 8005 & 2015-12-01 &Werner\\
        &  FUSE      & A0130202000&         & 20000      & 915--1188            & 4355 & 2000-11-23 &Dreizler\\
\noalign{\smallskip}
\pgsieben& HST/COS   & LCK203010 & G130M    &13000--19000& 1070--1210, 1224--1365&14344 & 2015-09-14 &Werner\\ 
        & HST/GHRS   & Z2T20304T & G140L    & 1800--2200 & 1165--1460           & 1197 & 1995-08-06 &Heber\\
        &  FUSE      & P1320401000&         & 20000      & 915--1188            &14541 & 2000-06-06 &Kruk\\

\noalign{\smallskip} \hline
\end{tabular} 
\tablefoot{\tablefoottext{a}{For the HST/COS observations the two wavelength
    intervals covered by detector segments A and B are
    given. All FUSE spectra were obtained with the LWRS
    aperture. Resolving power is $R$. Exposure time is $t_{\rm exp}$.}} 
\end{center}
\end{table*}

\section{Introduction}
\label{sect:Introduction}

The hot white dwarf \re is unique because a large number of trans-iron
elements were discovered in its photosphere
\citep{Werner_et_al_2012b}. Abundance analyses of 14 species with
atomic numbers in the range $Z=30-56$, i.e., zinc through barium,
revealed high quantities of up to about $10^4$ times the solar values
\citep[][and references therein]{rauch17b}. 

\re is a non-DA white dwarf (WD) with a helium-dominated atmosphere
\citep[spectral type hot DO, effective temperature \Teff $= 70\,000
  \pm 2000$\,K, surface gravity \logg $=7.5 \pm
  0.1$;][]{Rauch_et_al_2016b}. We concluded that the high heavy-metal
abundances are probably caused by radiative levitation
\citep{rauchetal2016mo}. If true, then we should find the same
phenomenon among other hot DOs with similar temperature and
gravity. Therefore we performed ultraviolet (UV) spectroscopy of
three other WDs with the Cosmic Origins Spectrograph (COS) aboard the
\emph{Hubble} Space Telescope (HST). Two of these WDs (\pgnull and \wdnull, alias
HS\,0111$+$0012) are also of the hot DO spectral type. The third
(\pgsieben) was drawn from the PG\,1159-type class, which are hot,
non-DA (pre-) WDs with large amounts of atmospheric carbon and
oxygen. Their surface chemistry is the result of a late He-shell flash
that dredged up the intershell matter located between the H and He
burning shells of the previous Asymptotic Giant Branch (AGB)
evolutionary phase \citep{Werner_Herwig_2006}. PG\,1159 stars without
remaining hydrogen evolve into DO WDs as soon as gravitational
settling starts to remove these species from the atmosphere. The
choice of a PG\,1159-type target was motivated by the fact that the
trans-iron element abundances are not yet affected by diffusion and,
hence, should reflect the abundances in the intershell layer created
by the s-process during the AGB phase.

This paper is structured as follows. First, we introduce our program
stars in Sect.\,\ref{sect:Stars}, and proceed with a description of
our new and archival observations
(Sect.\,\ref{sect:Observations}). We then present our spectral
analysis to determine the basic atmospheric parameters \Teff,
\logg, and metal abundances (Sect.\,\ref{sect:Analysis}). We discuss
our results and conclude in Sect.\,\ref{sect:Results_Conclusions}.

\section{The program stars}
\label{sect:Stars}

\begin{figure}
\resizebox{\hsize}{!}{\includegraphics{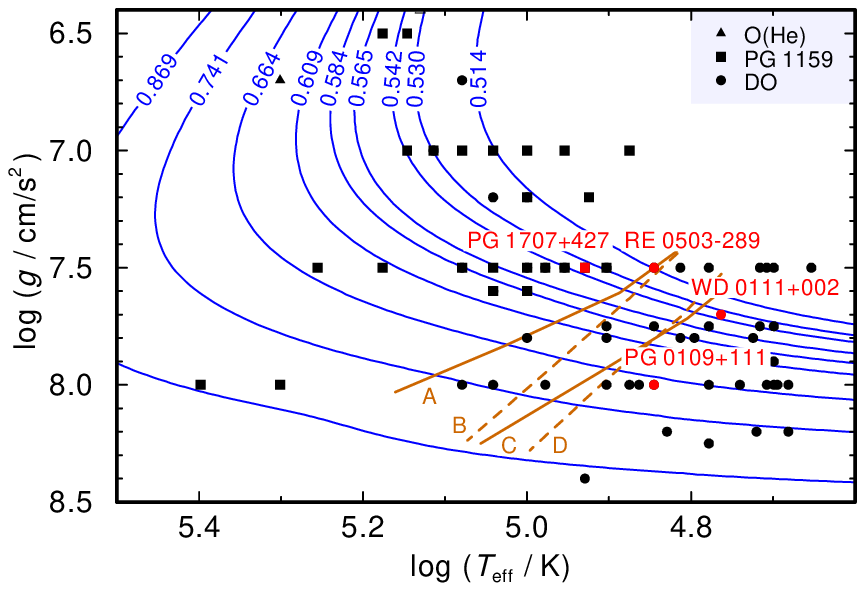}}
\caption{Location of our three program stars and the comparison star
  \re (red symbols) together with related objects
  \citep{Huegelmeyer_et_al_2006,Kepler_et_al_2016,Reindl_et_al_2014a,Werner_Herwig_2006}
  in the $\Teff - g$ plane. The single triangle symbol denotes a
    helium-dominated object of the O(He) spectral class
    \citep{Reindl_et_al_2014b}. Evolutionary tracks for H-deficient
  WDs \citep{Althaus_et_al_2009} labeled with the respective masses in
  \Msol are plotted for comparison. Four variants of the PG\,1159--DO
  transition limit \citep[labeled A--D,][]{Unglaub_Bues_2000} are
  indicated.}
\label{fig:VradVgred:tg_db_pg_pos}
\end{figure}

For our new observations we chose two hot DOs and a PG1159 star, which
are bright enough to obtain UV spectra with a good signal-to-noise ratio
enabling the identification of even weak metal lines. The location of
the program stars in the $g$--\Teff diagram (already with our
improved parameters for \pgnull) together with other hydrogen-deficient
stars is shown in Fig.\,\ref{fig:VradVgred:tg_db_pg_pos}.

\subsection{\wdnull and \pgnull (DO WDs)}

The non-local thermodynamic equilibrium (NLTE) analysis of optical
spectroscopy yielded \Teff $=65\,000 \pm 5000$\,K, \logg $= 7.8 \pm
0.3$ for \wdnull and \Teff $=110\,000 \pm 10\,000$\,K, \logg $=8.0 \pm
0.3$ for \pgnull. A carbon abundance of C=0.003 in both stars was derived
\citep{Dreizler_Werner_1996}; all abundances given
in this paper are mass fractions. \cite{Dreizler_1999} used HST/GHRS
spectra to assess metal abundances in \wdnull. With the exception of carbon lines,
only a few weak \ion{Ni}{v} lines were detected and an upper limit of
Ni = $1.4 \times 10^{-5}$ was established. Archival FUSE spectra of
both stars remained hitherto not analyzed.

\subsection{\pgsieben (PG1159 star)}

This star is a well-studied object \citep[][and references
  therein]{Werner_et_al_2015}. It has \Teff $=85\,000 \pm 5000$\,K and
\logg $=7.5 \pm 0.5$, and is dominated by helium and carbon in roughly
equal amounts. In the work cited, we assessed
the abundances of metals up to the iron group using far-UV spectra
taken with the Far Ultraviolet Spectroscopic Explorer (FUSE) and HST
spectra of relatively low resolution taken with the Goddard High
Resolution Spectrograph (GHRS). Our new HST/COS spectroscopy aimed at
spectra of better quality to detect weaker lines, particularly of
heavy metals.

\section{Observations}
\label{sect:Observations}

In Table\,\ref{tab:obs}, we list the observation log of our new HST/COS
data together with other archival UV HST and FUSE spectra used in
our study. The datasets were retrieved from the Barbara A. Mikulski
Archive for Space Telescopes (MAST). 
According to the COS Instrument Science Report
2009-01(v1)\footnote{\url{http://www.stsci.edu/hst/cos/documents/isrs/ISR2009_01.pdf}},
the line spread function (LSF) of the COS spectra deviates from a
Gaussian shape. We therefore convolve all synthetic spectra with the
correct LSF.

We used optical observations to constrain the surface gravity from
the ionized helium lines. Optical spectra for \wdnull\ and
\pgsieben\ were obtained from the Sloan Digital Sky Survey (SDSS;
observation IDs 1237663784742354959 and 1237665571981623391,
respectively). We also used the spectra of \pgnull from
\cite{Dreizler_Werner_1996} and \pgsieben from
\cite{1991A&A...244..437W}.

\subsection{Radial velocities}
\label{sect:VradVgred}

To shift the observations to rest wavelength, we determined the
stellar radial velocities $\vrad$ from the HST/COS spectra. To measure the
observed wavelengths by Gaussian fits, we employed IRAF\footnote{IRAF
  is distributed by the National Optical Astronomy  Observatory, which
  is operated by the Associated Universities for Research  in
  Astronomy, Inc., under cooperative agreement with the National
  Science Foundation.}.  The average values are  $\vrad =  44.8 \pm
5.4$\,km/s,  $\vrad =  37.2 \pm 4.7$\,km/s, and  $\vrad = -33.8 \pm
5.9$\,km/s for  \wdnull (from 11 lines), \pgnull (11 lines), and
\pgsieben (32 lines), respectively.
Considering error limits, our value for \wdnull is marginally in
agreement with wavelength measurements of several lines in HST/GHRS
spectra performed by \cite{Dreizler_1999}.

\begin{figure*}
\centering
\resizebox{0.68\textwidth}{!}{\includegraphics{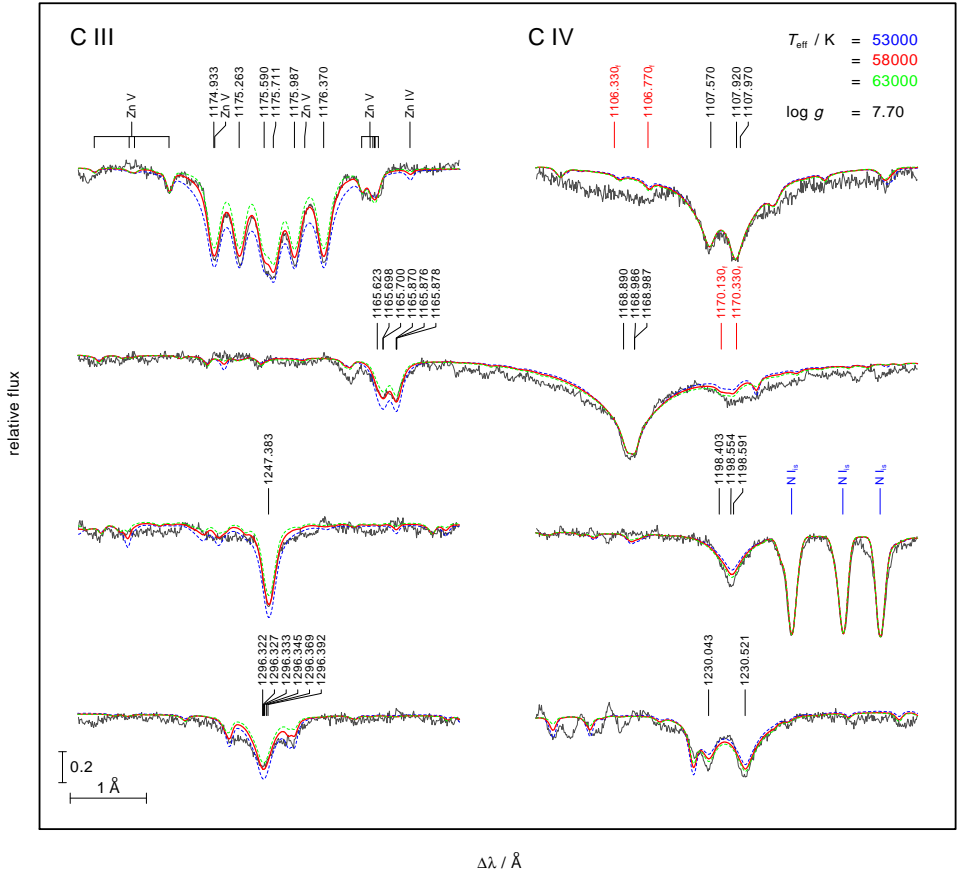}}
\caption{\ion{C}{iii} and \ion{C}{iv} lines in the HST/COS spectrum of
  \wdnull\ (black) compared  with three \logg = 7.7 models with
  different \Teff: 53\,000\,K (blue dashed), 58\,000\,K (red), and
  63\,000\,K (green dashed). Red wavelength labels indicate forbidden \ion{C}{iv}
  line components \citep{2016ApJ...827L...4W}. Also indicated are Zn lines near the
  \ion{C}{iii} multiplet at 1175\,\AA. The vertical bar indicates 20\% of the continuum flux.}
\label{fig:WD0111+002_Teff}
\end{figure*}

\subsection{Interstellar neutral hydrogen and reddening}
\label{sect:H_Column_Density}

We used the $\mathrm{Ly}\,\beta$ line profile in the FUSE spectra to
measure the interstellar neutral hydrogen column density $N_H$
(Fig.\,\ref{fig:H_Column_Density:Hbeta}). The results are
$1.1^{+0.8}_{-0.6} \times 10^{20}\,\mathrm{cm}^{-2}$ for \wdnull,
$1.8^{+0.8}_{-0.8} \times 10^{20}\,\mathrm{cm}^{-2}$ for \pgnull, and
$1.2^{+0.8}_{-0.6} \times 10^{20}\,\mathrm{cm}^{-2}$ for \pgsieben.
\cite{Dreizler_Heber_1998} used the $\mathrm{Ly}\,\alpha$ line in the
HST/GHRS spectrum of \pgsieben and found $1.5 \times
10^{20}\,\mathrm{cm}^{-2}$, that is consistent with our value
within its error limit.

We converted $N_H$ into reddening $E_\mathrm{B-V}$ using the empirical
relationship derived by \citet{Groenewegen_Lamers_1989}:  $\log
(N_H/E_\mathrm{B-V}) = 21.58 \pm 0.10$. We obtained $E_\mathrm{B-V} =
0.029^{+0.034}_{-0.018}$ for \wdnull, $0.047^{+0.039}_{-0.026}$ for
\pgnull, and $0.032^{+0.035}_{-0.019}$ for \pgsieben. In
Fig.\,\ref{fig:H_Column_Density:EBV}, we show our final models
attenuated with these reddening values together with the observed
continuum shape. The agreement is very good. To further improve the
fits, marginally lower reddening values (with error limits) for
\wdnull and \pgsieben are required. Our result for \pgsieben agrees
with the value of 0.02 derived by \cite{Dreizler_Heber_1998} from the
continuum slope of the GHRS spectrum.

\subsection{Unidentified lines}

There are many unidentified photospheric lines in the HST/COS spectra
of our program stars. Particularly conspicuous are features that are
seen in both our spectrum of \pgnull and in a spectrum of the
comparison star \re, taken with the Space Telescope Imaging
Spectrograph (STIS) aboard HST and presented by
\citet{2017A&A...598A.135H}; see Fig.\,\ref{fig:all_spectra}. These  features
potentially stem from trans-iron elements and we therefore list their
wavelength positions in Tab.\,\ref{tab:unid_lines}.

\begin{table}[t]
\begin{center}
\caption{Number of NLTE levels and lines in our model ions.\tablefootmark{a}}
\label{tab:modelatoms} 
\tiny
\begin{tabular}{clllllll}
\hline 
\hline 
\noalign{\smallskip}
   & I      & II      & III    & IV     & V      & VI    & VII  \\
\hline 
\noalign{\smallskip}
He & 19,35 & 16,120   &        &        &        &        \\
C  &        &         & 51,217 & 54,295 &        &        \\
N  &        &         & 34,129 & 90,546 & 54,297 &        \\
O  &        &         & 72,322 & 83,637 & 50,235\tablefootmark{b} &        \\
F  &        &         &        & 30,70  & 30,104 & 30,93\tablefootmark{c} \\
Ne &        &         & 46,126 & 40,126 & 94,641\tablefootmark{d} &        \\
Si &        &         & 17,28  & 30,102 & 25,59  &        \\
P  &        &         &        & 36,28  & 18,49  &        \\
S  &        &         &        & 57,332 & 39,107 & 25,48 \\
As &        &         &        & 14,7   &  5,10  & 14,5 & 7,1 \\
Sn &        &         & 21,9   & 10,16  &  9,5   & 15,5 &      \\
\noalign{\smallskip} \hline
\end{tabular} 
\tablefoot{ \tablefoottext{a}{First and second number of each table
    entry, separated by commas, denote the number of levels and lines,
    respectively. Not listed for each element is the highest
    considered ionization stage, which comprises its ground state
    only. See text for the treatment of iron-group and trans-iron
    elements besides As and Sn.}   
\tablefoottext{b}{For \wdnull,
    \ion{O}{v} was the highest ionization stage with one level.}
  \tablefoottext{c}{Only considered for \pgsieben.}
  \tablefoottext{d}{For \wdnull, \ion{Ne}{v} was the highest
    ionization stage with one level.}  } 
\end{center}
\end{table}

\begin{figure}
\centering
\resizebox{0.8\columnwidth}{!}{\includegraphics{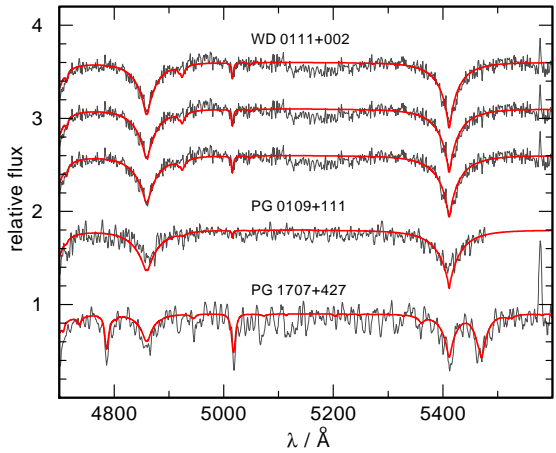}}
\caption{Observed spectra (black) around two lines of the \ion{He}{ii}
  Pickering series compared with synthetic spectra (red). From
  \emph{top} to \emph{bottom}: \wdnull with three models
  (\Teff$=58\,000$\,K and \logg = 7.6, 7.7, 7.8), \pgnull with
  model 70\,000/7.5, \pgsieben with model
  85\,000/7.5. The observations of
  \pgnull and \pgsieben were smoothed with a low-pass
  filter \citep{savitzkygolay1964}.  }
\label{fig:TeffLogg:Logg}
\end{figure}

\begin{figure}
\centering
\resizebox{0.8\columnwidth}{!}{\includegraphics{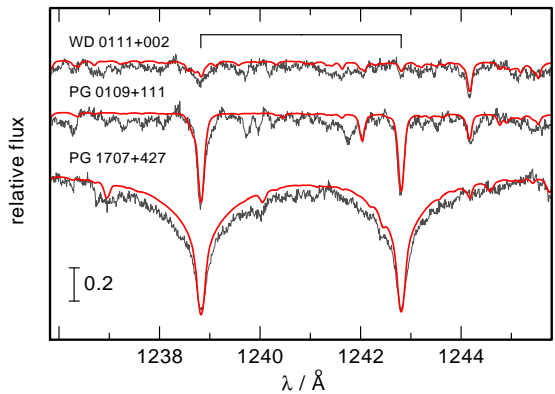}}
\caption{ \ion{N}{v} resonance doublet in the program stars (black) compared to our final models (red).
}
\label{fig:Element_Abundances:N}
\end{figure}

\begin{figure}
\centering
\resizebox{0.8\columnwidth}{!}{\includegraphics{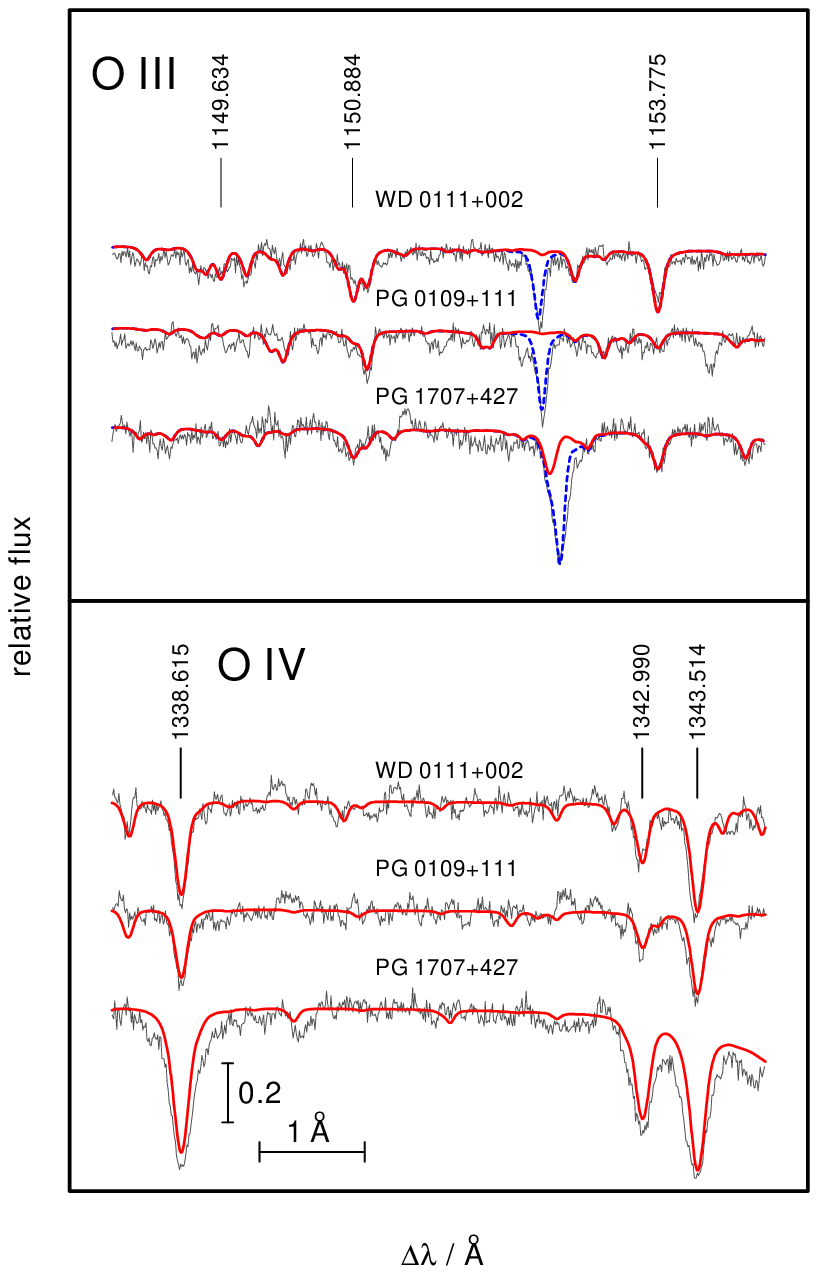}}
\caption{Like Fig.\,\ref{fig:Element_Abundances:N}, but for \ion{O}{iii} and
  \ion{O}{iv} lines.}
\label{fig:Element_Abundances:O}
\end{figure}

\section{Spectral analysis}
\label{sect:Analysis}

\subsection{Model atmospheres and atomic data}
\label{sect:Method}

To calculate chemically homogeneous, NLTE model atmospheres, we used
the T\"ubingen model-atmosphere package
\citep[TMAP\footnote{\url{http://astro.uni-tuebingen.de/~TMAP
 }};][]{Werner_et_al_2003,Werner_et_al_2012a}.  It assumes
plane-parallel geometry and radiative and hydrostatic
equilibrium. Atomic data were compiled from the T\"ubingen Model-Atom
Database \citep[TMAD;][]{Rauch_Deetjen_2003} that has been constructed
as a part of the T\"ubingen contribution to the German Astrophysical
Virtual Observatory (GAVO\footnote{\url{http://www.g-vo.org}}). A
summary of the model atoms for the included light elements up to
atomic number $Z=16$; As and Sn are given in
Tab.\,\ref{tab:modelatoms}. 

\begin{figure*}
\centering
\resizebox{0.68\textwidth}{!}{\includegraphics{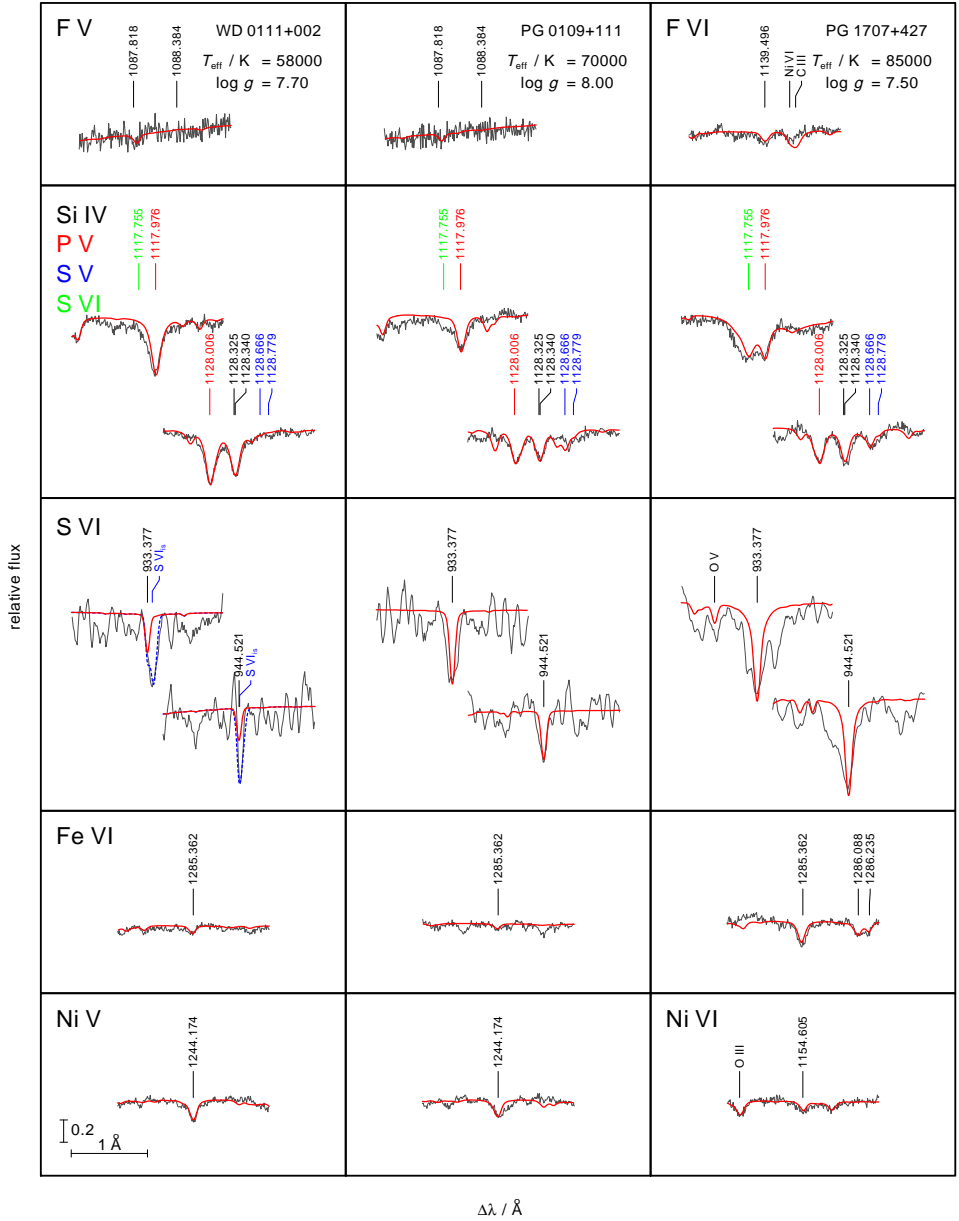}}
\caption{Synthetic line profiles for F, Si, P, S, Fe, and Ni in our
  final models (red) compared to observations (black) of our three
  program stars.}
\label{fig:Element_Abundances:F_Si_P_S_Fe_Ni}
\end{figure*}

\begin{figure*}
\centering
\resizebox{0.68\textwidth}{!}{\includegraphics{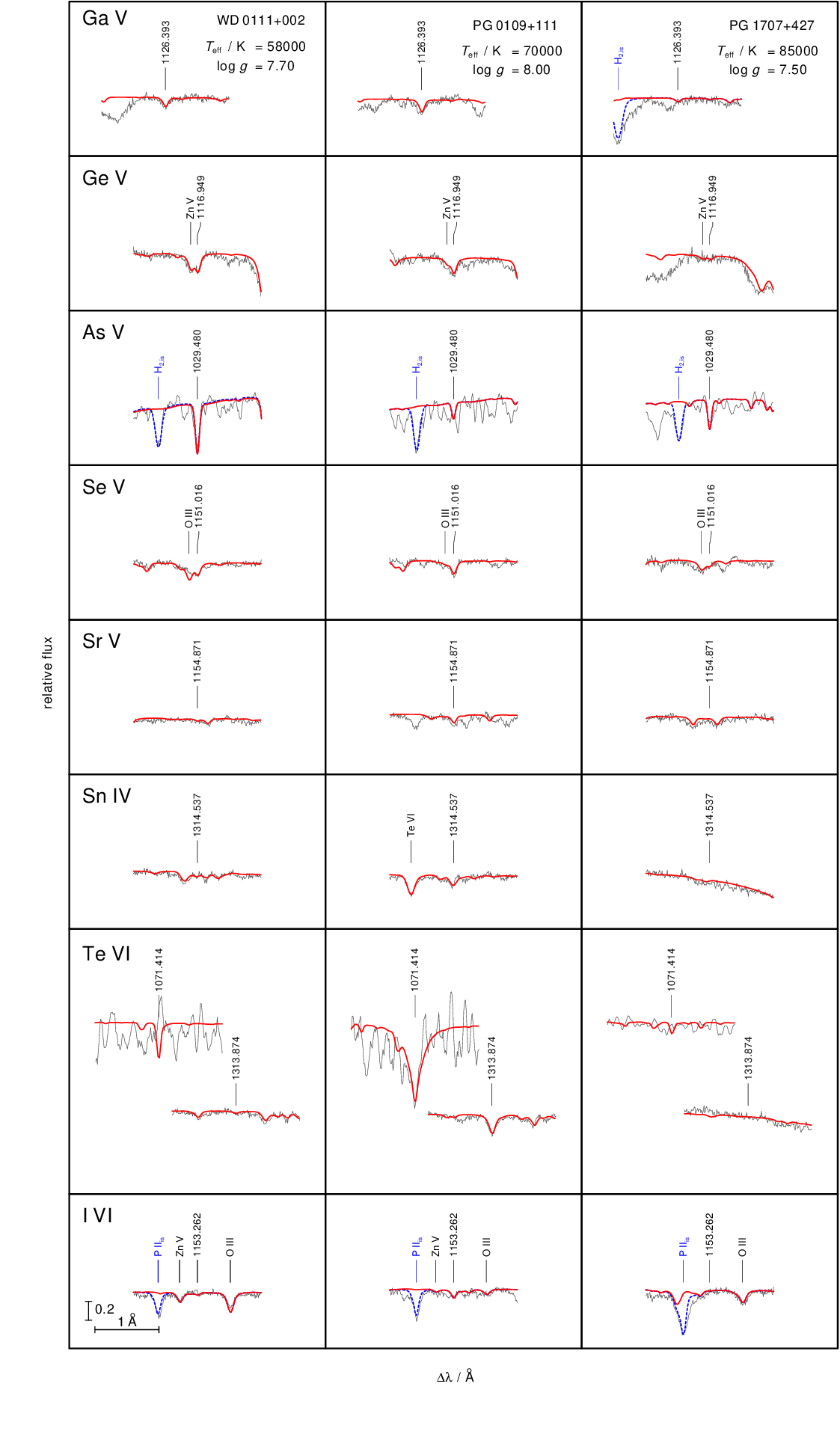}}
\caption{Like Fig.\,\ref{fig:Element_Abundances:F_Si_P_S_Fe_Ni}, but for Ga, Ge, As, Se, Sr, Sn, Te, and I.}
\label{fig:Element_Abundances:GaGeKrSr}
\end{figure*}

For iron and nickel, we used a statistical approach \citep[employing
  the IrOnIc tool;][]{Rauch_Deetjen_2003} with seven
superlevels per ion linked by superlines, together with an opacity
sampling method. Ionization stages \ion{Fe}{iii--vi} (plus
\ion{Fe}{vii} for the \pgsieben models) and \ion{Ni}{iii--vi}
augmented by a single, ground-level stage were considered per species
using the line lists of Kurucz
\citep{kurucz1991,kurucz2009,kurucz2011}. 

Motivated by the discovery of trans-iron elements in the DO \re
\citep{Werner_et_al_2012b}, respective transition probabilities have been calculated (for references see Table\,\ref{tab:tee}) and are
provided by the T\"ubingen Oscillator Strength Service (TOSS), which
is a part of the GAVO project. For our analysis, these species were
treated with the same statistical method as Fe and Ni \citep[][except for As
and Sn, for which classical model atoms as for the other species were
used]{Rauch_et_al_2015a}, considering lines of \ion{Zn}{iv--v},
\ion{Ga}{iv--vi}, \ion{Ge}{iv--vi}, \ion{As}{iv--vii},
\ion{Se}{iv--vi}, \ion{Kr}{iii--vii}, \ion{Sr}{iii--vii},
\ion{Zr}{iii--vii}, \ion{Mo}{iii--vii}, \ion{Sn}{iii--vi},
\ion{Te}{iv--vi}, \ion{I}{iv--vi}, \ion{Xe}{iii--vii}, and
\ion{Ba}{v--vii}.

The model atmospheres for all three program stars include He, C, N, O,
and Ne. The neon abundance cannot be determined, so we assumed the
following values: Ne = 0.01 for \pgsieben, which is a value that is typical for
PG\,1159 stars \citep{2004A&A...427..685W}, and rather small values
for \wdnull ($2.0\times10^{-5}$) and \pgnull ($1.5\times10^{-6}$). For
the two hottest program stars (\pgnull and \pgsieben), we also included
the following species into the model atmosphere calculations: F, Si,
P, S, Zn, Ga, and Ge. To save computing time, smaller versions of
the model atoms for all metals up to sulfur were used for the model
atmosphere calculations. Subsequently, the large versions (whose
numbers of levels and lines are listed in Table~\ref{tab:modelatoms})
were used to compute improved level occupation numbers while keeping
fixed the model-atmosphere structure. All other species heavier than
Ge were considered as trace elements, i.e., their backreaction on the
model structure was neglected.

\subsection{Effective temperature and surface gravity}
\label{sect:TeffLogg}

\noindent To measure \Teff, we evaluated the \Ion{C}{3}/\Ion{C}{4}
ionization equilibrium using the multiplets of \Ion{C}{3} at 1166,
1296, 1175\,\AA\ plus a singlet at 1247\,\AA, and the
multiplets of \Ion{C}{4} at 1108, 1169, 1198,
1230\,\AA\ (Figs.\,\ref{fig:WD0111+002_Teff},
\ref{fig:PG0109+111_Teff}, and \ref{fig:PG1707+427_Teff}). For the
determination of $g$ we used lines of the \Ion{He}{2} Pickering
series (Fig.\,\ref{fig:TeffLogg:Logg}). Various other ionization
equilibria (\Ion{N}{3}/\Ion{N}{4}/\Ion{N}{5},
\Ion{O}{3}/\Ion{O}{4}/\Ion{O}{5}/\Ion{O}{6}, \Ion{F}{5}/\Ion{F}{6},
\Ion{S}{5}/\Ion{S}{6}, \Ion{Fe}{5}/\Ion{Fe}{6}/\Ion{Fe}{7},
\Ion{Zn}{4}/\Ion{Zn}{5} \Ion{Ge}{4}/\Ion{Ge}{5},
\Ion{Ga}{4}/\Ion{Ga}{5}; not all balances available in all stars)
confirm the results for \Teff\ and \logg\ found by this procedure.

For \wdnull, we found \Teff = $58\,000\pm5000$\,K and \logg =
$7.7\pm0.3$, in agreement with previous results obtained from
optical spectra alone \citep[65\,000/7.8; ][]{Dreizler_Werner_1996}.
In the case of \pgnull, we derived \Teff = $70\,000\pm5000$\,K and
\logg = $8.0\pm0.3$. In comparison to previous results
\citep[110\,000/8.0; ][]{Dreizler_Werner_1996}, the temperature is
significantly lower. In the work cited, it was noted that indeed the
weak, observed \Ion{He}{1} lines are not compatible with the
high-temperature model that displays no neutral helium lines at all.
For \pgsieben our results (\Teff = $85\,000\pm5000$\,K, \logg =
$7.5\pm0.3$) confirm previous analyses of UV and optical data
\citep[][and references therein]{Werner_et_al_2015}.

\begin{figure*}
\centering
\resizebox{0.9\textwidth}{!}{\includegraphics{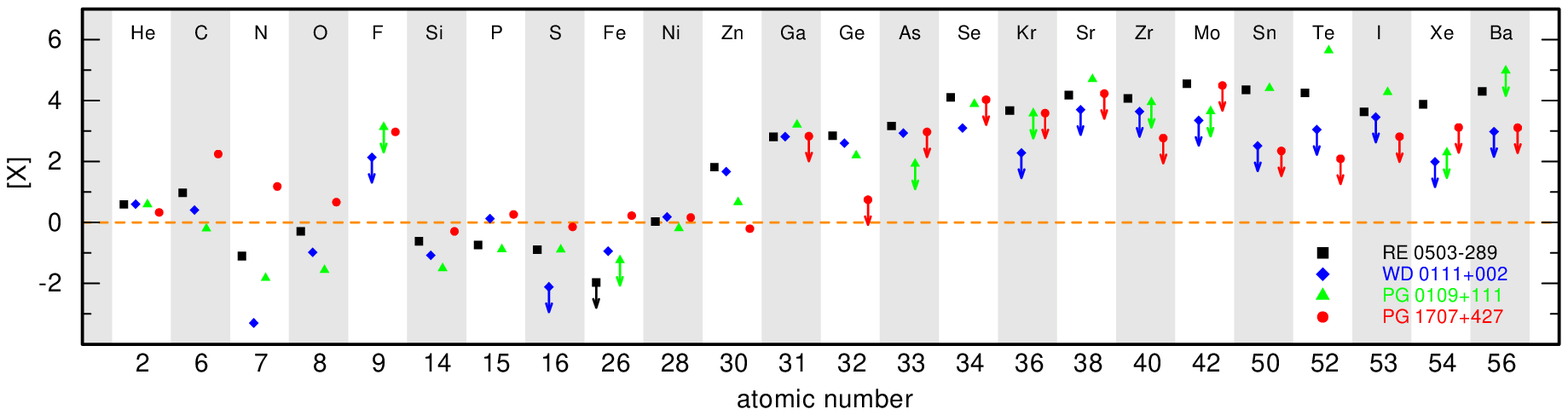}}
\caption{Photospheric abundances of our three program stars and the
  comparison star \re. [X] denotes $\log(\mathrm{abundance} /
  \mathrm{solar\ abundance})$ of species X. The dashed horizontal line
  indicates solar abundances. }
\label{fig:Results:Abundances}
\end{figure*}

\subsection{Element abundances}
\label{sect:Element_Abundances}

Element abundances were derived by detailed line-profile fits. Upper limits
were established by reducing the abundances such that computed
spectral lines become undetectable regarding the signal-to-noise
ratio. We describe the procedure for the different species and display
some representative fits. Results are given in Table\,\ref{tab:abundances}
and Fig.\,\ref{fig:Results:Abundances}. Typical errors are estimated
to 0.3\,dex. A comparison of the complete HST/COS spectra with the final
models appears in the Appendix (Fig.\,\ref{fig:all_spectra}).

\paragraph{Carbon, nitrogen, and oxygen.}

For carbon, we used the same lines as for the \Teff determination
(Figs.\,\ref{fig:WD0111+002_Teff}, \ref{fig:PG0109+111_Teff},
\ref{fig:PG1707+427_Teff}). For nitrogen, the \ion{N}{v}
resonance doublet is the primary
indicator \sA{fig:Element_Abundances:N}, and for oxygen the
\ion{O}{iii} $\lambda\lambda$1149.6, 1150.9, 1153.8\,\AA\ and
\ion{O}{iv} $\lambda\lambda$1338.6, 1343.0, 1343.5\,\AA\ lines
\sA{fig:Element_Abundances:O}.

\paragraph{Fluorine, silicon, phosphorus, and sulfur.}

For these species, we proceeded as described in detail for the
analysis of the FUSE spectrum of \pgsieben
\citep{Werner_et_al_2015}. The following lines were used: \ion{F}{v}
$\lambda\lambda$1087.8, 1088.4\,\AA, \ion{F}{vi} $\lambda$1139.5\,\AA,
\ion{Si}{iv} $\lambda\lambda$1066.6, 1122.5, 1128.3\,\AA, \ion{P}{v}
$\lambda\lambda$1000.4, 1118.0, 1128.0\,\AA, \ion{S}{v}
$\lambda\lambda$1128.7, 1128.8\,\AA, \ion{S}{vi}
$\lambda\lambda$933.4, 944.5, 1117.8\,\AA. The fits to most of these lines
are presented in Fig.\,\ref{fig:Element_Abundances:F_Si_P_S_Fe_Ni}.

\paragraph{Iron and nickel.}

Iron could not be detected in \pgnull, but a few weak \ion{Fe}{vi}
lines are detected in \wdnull. In the COS spectra of \pgsieben, we see many lines
from \ion{Fe}{vi} and few from \ion{Fe}{vii}. The identification of
iron was uncertain in the FUSE spectra \citep{Werner_et_al_2015}. We
detect many nickel lines (\ion{Ni}{v}) in the COS spectra of \wdnull
and \pgnull. In \pgsieben, the presence of \ion{Ni}{vi} lines in FUSE
spectra is uncertain \citep{Werner_et_al_2015}, but in the COS
spectra we detect two lines at 1124.19 and
1154.60\,\AA\ (Fig.\,\ref{fig:Element_Abundances:F_Si_P_S_Fe_Ni},
Tables~\ref{tab:Element_Abundances:Iron1},
\ref{tab:Element_Abundances:Iron2}, and \ref{tab:nickel}). 

\paragraph{Zinc.}

Many \ion{Zn}{v} lines are present in the COS data of our program
stars; these lines are most prominent in \wdnull, followed by \pgnull, and are rather weak
in \pgsieben. Among the strongest features is a \ion{Zn}{v} blend at
1177\,\AA\ adjacent to the \Ion{C}{3} $\lambda$\,1175\,\AA\ multiplet
(Figs.\,\ref{fig:WD0111+002_Teff}, \ref{fig:PG0109+111_Teff},
\ref{fig:PG1707+427_Teff}). \wdnull also shows \ion{Zn}{iv} lines,
for example, at 1349.88\,\AA.

\paragraph{Gallium and germanium.}

No lines are detectable in \pgsieben, but we see a number of
\ion{Ga}{iv} and \ion{Ga}{v} as well as \ion{Ge}{iv} and \ion{Ge}{v}
lines in \wdnull and \pgnull
(Fig.\,\ref{fig:Element_Abundances:GaGeKrSr}).

\paragraph{As, Se, Sr Sn, Te, and I.}

We used the \ion{As}{v} $\lambda\lambda$987.7/1029.48\,\AA\ resonance
doublet in the FUSE spectra to determine the As abundance in
\wdnull\ and upper limits for the other two stars. In the case of
\pgsieben, ISM lines are located at the wavelength position of the
\ion{As}{v} doublet \citep[Fig.\,A.1 in][]{Werner_et_al_2015}, such
that we can derive an upper limit only. \ion{Se}{v}
$\lambda\lambda$1151.0, 1227.5\,\AA\ can be seen in both DOs but not
in \pgsieben. Lines from Sr, Sn, Te, and I are only seen in
\pgnull\ (\ion{Sr}{v} $\lambda$1154.9\,\AA, \ion{Sn}{iv}
$\lambda$1314.5\,\AA, \ion{Te}{vi} $\lambda$1313.9\,\AA, \ion{I}{vi}
$\lambda$1153.3\,\AA) but not in the other two stars. For line fits,
see Fig.\,\ref{fig:Element_Abundances:GaGeKrSr}.

In \pgnull, the fitted \ion{Sn}{iv} $\lambda$1314.5\,\AA\ line is
rather weak. A
number of detected \ion{Sn}{v} lines would be better suited for a more precise analysis, but oscillator strengths are
unavailable. \ion{Sn}{v} lines are not detected in \pgsieben,
but they would give a much tighter upper limit for the Sn abundance
than that derived from the absence of the \ion{Sn}{iv} line.

The tellurium abundance in \pgnull is extremely high (5.6 dex
oversolar). Its mass fraction of $6.2\times10^{-3}$ means that it is
the most abundant metal in the atmosphere. The abundance was derived
from the \ion{Te}{vi} $\lambda$1313.9\,\AA\ line (see
Fig.\,\ref{fig:Element_Abundances:GaGeKrSr}). In fact, there are two
much stronger \ion{Te}{vi} lines, namely the resonance doublet at
951.0 and 1071.4\,\AA, but they are located in rather noisy portions
of the FUSE and HST/COS spectra. Nevertheless, these lines can be detected in
the observations (see the 1071.4\,\AA\ component in
Fig.\,\ref{fig:all_spectra}) as very strong and broad features.
They are among the strongest metal line features rivaling the
prominent \ion{C}{iii} and \ion{C}{iv} lines.

\paragraph{Kr, Zr, Mo, Xe, and Ba.}

No lines of these species could be identified in our program stars.

\begin{table*}[t]
\begin{center}
\caption{Final adopted atmospheric parameters of the three program
  stars and the comparison star \re (columns 2--5). In the next two
  columns, trans-iron abundances from literature for two cool DOs
  (HD\,149499B and HZ\,21) are listed. The last column gives the solar
  abundances.\tablefootmark{a}}
\label{tab:abundances} 
\begin{tabular}{rrrrrrrr}
\hline 
\hline 
\noalign{\smallskip}
                     & WD\,0111            & PG\,0109            & PG\,1707            & RE\,0503            & HD                & HZ\,21              &  Sun \\
                     & $+$002              & $+$111              & $+$427              & $-$289              & 149499B \\
\noalign{\smallskip}
\hline 
\noalign{\smallskip}
Type                 &  DO                 & DO                  & PG1159              & DO                  & DOA               & DO\\
\Teff/\,K            & 58\,000             & 70\,000             & 85\,000             & 70\,000             & 49\,500           & 53\,000\\
\logg                & 7.7                 & 8.0                 & 7.5                 & 7.5                 & 7.97              & 7.8\\
\noalign{\smallskip}
He                   &  $9.9\times10^{-1}$ &  $9.9\times10^{-1}$ &  $5.3\times10^{-1}$ &  $9.7\times10^{-1}$ &                   &                    & $2.5\times10^{-1}$ \\
C                    &  $6.0\times10^{-3}$ &  $1.5\times10^{-3}$ &  $4.1\times10^{-1}$ &  $2.2\times10^{-2}$ &                   &                    & $2.4\times10^{-3}$ \\
N                    &  $3.5\times10^{-7}$ &  $1.0\times10^{-5}$ &  $1.0\times10^{-2}$ &  $5.5\times10^{-5}$ &                   &                    & $6.9\times10^{-4}$ \\
O                    &  $6.0\times10^{-4}$ &  $1.6\times10^{-4}$ &  $2.7\times10^{-2}$ &  $2.9\times10^{-3}$ &                   &                    & $5.7\times10^{-3}$ \\
F                    & $<4.7\times10^{-5}$ & $<4.7\times10^{-4}$ &  $3.2\times10^{-4}$ &                     &                   &                    & $3.5\times10^{-7}$ \\
Si                   &  $5.6\times10^{-5}$ &  $2.1\times10^{-5}$ &  $3.4\times10^{-4}$ &  $1.6\times10^{-4}$ &                   &                    & $6.7\times10^{-4}$ \\
P                    &  $7.7\times10^{-6}$ &  $7.6\times10^{-7}$ &  $1.1\times10^{-5}$ &  $1.1\times10^{-6}$ &                   &                    & $5.8\times10^{-6}$ \\
S                    & $<2.4\times10^{-6}$ &  $4.0\times10^{-5}$ &  $2.2\times10^{-4}$ &  $4.0\times10^{-5}$ &                   &                    & $3.1\times10^{-4}$ \\
Fe                   &  $1.4\times10^{-4}$ & $<6.9\times10^{-5}$ &  $2.0\times10^{-3}$ & $<1.3\times10^{-5}$ &                   &                    & $1.2\times10^{-3}$ \\
Ni                   &  $1.0\times10^{-4}$ &  $4.3\times10^{-5}$ &  $9.9\times10^{-5}$ &  $7.2\times10^{-5}$ &                   &                    & $6.8\times10^{-5}$ \\
\noalign{\smallskip}
Zn                   &  $8.2\times10^{-5}$ &  $8.1\times10^{-6}$ &  $1.1\times10^{-6}$ &  $1.1\times10^{-4}$ &                   &                    & $1.7\times10^{-6}$ \\
Ga                   &  $3.5\times10^{-5}$ &  $8.6\times10^{-5}$ & $<3.6\times10^{-5}$ &  $3.4\times10^{-5}$ &                   &                    & $5.3\times10^{-8}$ \\
Ge                   &  $9.1\times10^{-5}$ &  $3.6\times10^{-5}$ & $<1.3\times10^{-6}$ &  $1.6\times10^{-4}$ &                   &                    & $2.3\times10^{-7}$ \\
As                   &  $9.3\times10^{-6}$ & $<9.2\times10^{-7}$ & $<1.0\times10^{-5}$ &  $1.6\times10^{-5}$ & $8.1\times10^{-6}$ &  $3.0\times10^{-7}$ & $1.1\times10^{-8}$ \\
Se                   &  $1.6\times10^{-4}$ &  $9.6\times10^{-4}$ & $<1.3\times10^{-3}$ &  $1.6\times10^{-3}$ & $1.3\times10^{-5}$ & $<7.5\times10^{-6}$ & $1.3\times10^{-7}$ \\
Kr                   & $<2.1\times10^{-5}$ & $<4.1\times10^{-4}$ & $<4.2\times10^{-4}$ &  $5.0\times10^{-4}$ &                   &                    & $1.1\times10^{-7}$ \\
Sr                   & $<2.2\times10^{-4}$ &  $2.2\times10^{-3}$ & $<7.3\times10^{-4}$ &  $6.5\times10^{-4}$ &                   &                    & $4.3\times10^{-8}$ \\
Zr                   & $<1.1\times10^{-4}$ & $<2.3\times10^{-4}$ & $<1.5\times10^{-5}$ &  $3.0\times10^{-4}$ &                   &                    & $2.6\times10^{-8}$ \\
Mo                   & $<1.2\times10^{-5}$ & $<2.4\times10^{-5}$ & $<1.7\times10^{-4}$ &  $1.9\times10^{-4}$ &                   &                    & $5.3\times10^{-9}$ \\
Sn                   & $<2.9\times10^{-6}$ &  $2.3\times10^{-4}$ & $<2.0\times10^{-6}$ &  $2.0\times10^{-4}$ & $8.3\times10^{-6}$ &                    & $9.1\times10^{-9}$ \\
Te                   & $<1.6\times10^{-5}$ &  $6.2\times10^{-3}$ & $<1.7\times10^{-6}$ &  $2.5\times10^{-4}$ & $9.1\times10^{-7}$ &  $1.3\times10^{-6}$ & $1.4\times10^{-8}$ \\
I                    & $<9.4\times10^{-6}$ &  $6.2\times10^{-5}$ & $<2.2\times10^{-6}$ &  $1.4\times10^{-5}$ & $1.9\times10^{-5}$ & $<1.2\times10^{-6}$ & $3.3\times10^{-9}$ \\
Xe                   & $<1.6\times10^{-6}$ & $<3.2\times10^{-6}$ & $<2.2\times10^{-5}$ &  $1.3\times10^{-4}$ &                   &                    & $1.7\times10^{-8}$ \\
Ba                   & $<1.7\times10^{-5}$ & $<1.7\times10^{-3}$ & $<2.3\times10^{-5}$ &  $3.6\times10^{-4}$ &                   &                    & $1.8\times10^{-8}$ \\
\hline
\end{tabular} 
\tablefoot{\tablefoottext{a}{Abundances in mass fractions (see also
    Fig.\,\ref{fig:Results:Abundances}) and surface gravity $g$ in
    cm\,s$^{-2}$. Parameters for \re from \cite{rauch17b} and
    references therein. Trans-iron abundances for HD\,149499B and HZ\,21
    from \citet{2005ApJ...630L.169C}. Solar abundances
    from
    \citet{2009ARA&A..47..481A,scottetal2015b,scottetal2015a,grevesseetal2015}.
}  } 
\end{center}
\end{table*}

\section{Summary and conclusions}
\label{sect:Results_Conclusions}

We have analyzed new UV spectra taken with HST/COS of one PG\,1159
star and two hot DO WDs. The primary aim of this study was the identification
and abundance determination of trans-iron group elements to reach a conclusion about the effects of s-process and element diffusion. To this end, we
computed line-blanketed NLTE model atmospheres and began with an
assessment of the effective temperatures and surface gravities. For
both DOs, the evaluation of metal ionization balances resulted in
improved \Teff determinations yielding lower values than found in previous
analyses of optical spectra. In particular, \Teff of \pgnull is
70\,000\,K instead of 110\,000\,K. 

\paragraph{\pgnull and \wdnull (DO WDs).}

For the two DOs, we measured the trace-element abundances and compare
these to the DO \re, a hitherto unique DO in which a total of 14
trans-iron group elements (i.e., atomic number $Z>28$) have been
identified, with extreme overabundances (about 2--4.5~dex) relative to
the solar values (Tab.\,\ref{tab:abundances},
Fig.\,\ref{fig:Results:Abundances}). Both investigated DOs also
exhibit trans-iron species, but there are fewer of these species (five and
eight in \wdnull and \pgnull, respectively). In both DOs we detect Zn,
Ga, Ge, and Se, plus the heavier species Sr, Sn, Te, and I only in
\pgnull, and As only in \wdnull. Generally, their abundances are
(within about 1 dex) similar to \re with the exception of Te. This
element is almost 6\,dex oversolar, significantly exceeding the
enrichment in \re.

The general tendency that \re (\Teff= 70\,000\,K, logg=7.5) has the
highest number of detected trans-iron species can be explained by the
fact that radiative acceleration is probably more effective compared
to \pgnull, which has a higher gravity (70\,000/8.0), and compared to
\wdnull (with the lowest number of detected trans-iron species), which
is significantly cooler and has a slightly higher surface gravity
(58\,000/7.7).

\citet{2005ApJ...630L.169C} detected trans-iron elements in two cool DO
WDs, namely HZ\,21 \citep[\Teff = 53\,000\,K, \logg =
  7.8;][]{Dreizler_Werner_1996} and HD\,149499B \citep[\Teff =
  49\,500\,K, \logg = 7.97;][]{1995A&A...300L...5N}. Their abundances
are also given in Tab.\,\ref{tab:abundances}. The numbers listed (mass
fractions) were computed from the number ratios reported in
\citet{2005ApJ...630L.169C} (averaged over results from several lines,
if applicable), accounting for the fact that HD\,149499B is a DOA WD
with 6\% hydrogen (by mass). Generally, the trans-iron element
abundances are lower than in the hot DOs, probably owing to weaker
levitation at lower temperatures. 

As to the lighter elements, these are also generally less abundant in
the two DOs investigated here compared to \re, with the exception of
P, that is 0.8~dex higher in \wdnull than in \re. A noteworthy
commonality of all three DOs is the low Fe/Ni ratio of $\leq$ 1.6, in
contrast to the solar ratio Fe/Ni = 18.

\paragraph{\pgsieben (PG\,1159 star).} Our results for \Teff and \logg
agree with those in \cite{Werner_et_al_2015}. The same holds,
within error limits, for the element abundances. For the first time,
however, we identify nickel lines and find a solar Ni abundance. The most
significant result from our new HST/COS spectroscopy concerns the
trans-iron elements. While only high upper abundance limits could be
derived previously, we now determined the Zn abundance and, for most
other species, were able to reduce the upper abundance limits significantly
compared to our previous work involving FUSE data alone. We find that
Zn is solar and Ge is at most 0.8\,dex oversolar, suggesting that there
is no general overabundance of trans-iron elements. The upper limits for
other trans-iron elements are still high, however, between about 2.8 and
4.5~dex oversolar. In comparison to the DOs, this confirms that
radiative levitation does not affect the element abundances in the
atmosphere of \pgsieben. This is a plausible result for a PG\,1159
star because, as outlined in the introduction, the star is located
before the wind limit so that diffusion cannot affect the element
abundances. In addition, from the helium and light metal
abundances we concluded to see intershell matter composition
\citep{Werner_et_al_2015}.

It is expected from stellar evolution models that, as
discussed in detail in \cite{Werner_et_al_2015} based on the models
of a 2\,$M_\odot$ star after the 30th thermal pulse \citep[Gallino,
  priv. comm.;][]{2007ApJ...656L..73K} s-process elements should
be overabundant. But our results do not allow a quantitative
comparison because the predicted enhancement of Zn and Ge to
$<0.4$\,dex oversolar values is near the error margin of our
analysis. Much stronger enhancements are predicted by stellar models
for other trans-iron elements, but they are still below our upper limits
derived. For example, we could reduce the upper detection limit for
germanium by 1.6\,dex down to Ge $<1.3\times10^{-6}$, but the predicted
value is Ge $<5.9\times10^{-7}$, i.e., just barely below our detection
threshold. The strongest enhancement among the investigated s-process
elements is predicted for barium, namely Ba
$=1.2\times10^{-5}$. Again, although we were able to reduce the detection
threshold by 1.8\,dex, the current limit of Ba $<2.3\times10^{-5}$ is
slightly smaller. Further progress can only be achieved with UV
spectra with even better signal-to-noise ratio.

\begin{acknowledgements}
DH was supported by the German Aerospace Center (DLR, grant
50\,OR\,1501). The authors acknowledge support by the High
Performance and Cloud Computing Group at the Zentrum f\"ur
Datenverarbeitung of the University of T\"ubingen, the state of
Baden-W\"urttemberg through bwHPC, and the German Research Foundation
(DFG) through grant no INST 37/935-1 FUGG.  The  TIRO
(\url{http://astro-uni-tuebingen.de/~TIRO}) TMAD
(\url{http://astro-uni-tuebingen.de/~TMAD}) and TOSS
(\url{http://astro-uni-tuebingen.de/~TOSS}) services used for this
paper  were constructed as part of the activities of the German
Astrophysical Virtual Observatory.  Some of the data presented in this
paper were obtained from the Mikulski Archive for Space Telescopes
(MAST). STScI is operated by the Association of Universities for
Research in Astronomy, Inc., under NASA contract NAS5-26555. Support
for MAST for non-HST data is provided by the NASA Office of Space
Science via grant NNX09AF08G and by other grants and contracts.  This
research has made use of NASA's Astrophysics Data System and the
SIMBAD database, operated at CDS, Strasbourg, France.
\end{acknowledgements}

\bibliographystyle{aa}
\bibliography{aa}


\appendix

\section{Additional figures and tables}
\label{A:Effective_Temperatures}

\begin{table}
\setlength{\tabcolsep}{0.3em}
\caption{Wavelengths (in \AA) of unidentified photospheric
  lines. These possibly stem from trans-iron elements and are present in both,
  the HST/COS spectrum of \pgnull and the HST/STIS spectrum of \re.}
\label{tab:unid_lines}
\tiny
\begin{center}
\begin{tabular}{llllllllll}
\hline
\hline
\noalign{\smallskip}
1102.15, 1103.28, 1103.47, 1104.20, 1114.10, 1119.20, 1120.85, 1122.14\\
1123.52, 1137.22, 1157.55, 1157.90, 1159.79, 1161.88, 1165.30, 1174.66\\
1178.61, 1201.28, 1201.53, 1234.32, 1239.72, 1241.78, 1247.71, 1250.30\\
1254.09, 1259.91, 1262.26, 1264.11, 1274.72, 1278.18, 1278.92, 1345.67\\
1346.16\\
\hline
\end{tabular}
\end{center}
\end{table}

\begin{table}
\setlength{\tabcolsep}{0.3em}
\caption{Ions of trans-iron elements with recently calculated oscillator strengths.}
\label{tab:tee}
\begin{center}
\begin{tabular}{rrll}
\hline
\hline
\noalign{\smallskip}
Zn & {\sc iv} &-\hspace{2mm}{\sc v}            & \citet{rauchetal2014zn} \\
Ga & {\sc iv} &-\hspace{2mm}{\sc vi}           & \citet{Rauch_et_al_2015a} \\
Ge & {\sc  v} &-\hspace{2mm}{\sc vi}           & \citet{rauchetal2012ge} \\
Se & {\sc  v} &                                & \citet{rauch17b} \\
Kr & {\sc iv} &-\hspace{2mm}{\sc vii}          & \citet{Rauch_et_al_2016b} \\
Sr & {\sc iv} &-\hspace{2mm}{\sc vii}          & \citet{rauch17b} \\
Zr & {\sc iv} &-\hspace{2mm}{\sc vii}          & \citet{rauchetal2016zr} \\
Mo & {\sc iv} &-\hspace{2mm}{\sc vii}          & \citet{rauchetal2016mo} \\
Te & {\sc vi} &                                & \citet{rauch17b} \\
I  & {\sc vi} &                                & \citet{rauch17b} \\
Xe & {\sc iv} &-\hspace{2mm}{\sc v}, {\sc vii} & \citet{rauchetal2015xe,rauchetal2016zr} \\
Ba & {\sc  v} &-\hspace{2mm}{\sc vii}          & \citet{rauchetal2014ba} \\
\hline
\end{tabular}
\end{center}
\end{table}

\begin{table}
        \caption{\ion{Fe}{v} lines identified in the HST/COS spectrum of \pgsieben.}
        \begin{center}
                \begin{tabular}{r@{.}lcccl}
                        \hline
                        \hline
                        \noalign{\smallskip}
                        \multicolumn{2}{c}{Wavelength\,/\,\AA} & \multirow{2}{*}{Ion} & \multicolumn{2}{c}{Transition} & \multirow{2}{*}{Comment} \\
                        \cline{1-2}
                        \cline{4-5}
                        \noalign{\smallskip}
                        \multicolumn{2}{c}{laboratory}               &                      & low & up                       & \\
                        \noalign{\smallskip}
                        \hline
                        \noalign{\smallskip}
      1320&409 & \Ion{Fe}{5} & 4s $^3$H$^\mathrm{ }_{4}$    & 4p $^3$G$^\mathrm{o}_{3}$    & weak \\
      1321&341 & \Ion{Fe}{5} & 4s $^3$G$^\mathrm{ }_{4}$    & 4p $^3$H$^\mathrm{o}_{5}$    & weak \\
      1321&489 & \Ion{Fe}{5} & 4s $^3$G$^\mathrm{ }_{5}$    & 4p $^3$H$^\mathrm{o}_{6}$    & weak \\
      1323&271 & \Ion{Fe}{5} & 4s $^3$H$^\mathrm{ }_{5}$    & 4p $^3$G$^\mathrm{o}_{4}$    & weak \\
      1330&405 & \Ion{Fe}{5} & 4s $^3$H$^\mathrm{ }_{6}$    & 4p $^3$G$^\mathrm{o}_{5}$    & \\
                        1331&639 & \Ion{Fe}{5} & 4s $^1$D$^\mathrm{ }_{2}$    & 4p $^1$?$^\mathrm{o}_{2}$    & \\
      1361&278 & \Ion{Fe}{5} & 4s $^3$P$^\mathrm{ }_{2}$    & 4p $^3$F$^\mathrm{o}_{3}$    & weak \\
      1361&446 & \Ion{Fe}{5} & 4s $^1$F$^\mathrm{ }_{3}$    & 4p $^1$G$^\mathrm{o}_{4}$    & weak \\
      1361&691 & \Ion{Fe}{5} & 4s $^1$D$^\mathrm{ }_{2}$    & 4p $^1$F$^\mathrm{o}_{3}$    & \\
      1361&826 & \Ion{Fe}{5} & 4s $^3$F$^\mathrm{ }_{4}$    & 4p $^3$G$^\mathrm{o}_{5}$    & \\
      1362&864 & \Ion{Fe}{5} & 4s $^3$D$^\mathrm{ }_{3}$    & 4p $^3$D$^\mathrm{o}_{3}$    & weak \\
      1363&076 & \Ion{Fe}{5} & 4d $^5$F$^\mathrm{ }_{3}$    & 4p $^5$F$^\mathrm{o}_{4}$    & weak \\
                        \noalign{\smallskip}
                        \hline
                \end{tabular}
        \end{center}
        \label{tab:Element_Abundances:Iron1}
\end{table}

\begin{table}
        \caption{\ion{Fe}{vi} lines identified in the HST/COS spectrum of \pgsieben.}
\tiny
        \begin{center}
                \begin{tabular}{r@{.}lcccl}
                        \hline
                        \hline
                        \noalign{\smallskip}
                        \multicolumn{2}{c}{Wavelength\,/\,\AA} & \multirow{2}{*}{Ion} & \multicolumn{2}{c}{Transition} & \multirow{2}{*}{Comment} \\
                        \cline{1-2}
                        \cline{4-5}
                        \noalign{\smallskip}
                        \multicolumn{2}{c}{laboratory}               &                      & low & up                       & \\
                        \noalign{\smallskip}
                        \hline
                        \noalign{\smallskip}
      1115&099 & \Ion{Fe}{6} & 4s $^2$D$^\mathrm{ }_{5/2}$  & 4p $^2$D$^\mathrm{o}_{5/2}$  & \\
      1120&933 & \Ion{Fe}{6} & 4s $^2$F$^\mathrm{ }_{7/2}$  & 4p $^2$F$^\mathrm{o}_{7/2}$  & \\
      1121&147 & \Ion{Fe}{6} & 4s $^2$F$^\mathrm{ }_{5/2}$  & 4p $^2$F$^\mathrm{o}_{5/2}$  & \\
      1152&771 & \Ion{Fe}{6} & 4s $^2$G$^\mathrm{ }_{7/2}$  & 4p $^2$F$^\mathrm{o}_{5/2}$  & blend \Ion{P}{2} i.s. \\
      1160&509 & \Ion{Fe}{6} & 4s $^2$P$^\mathrm{ }_{1/2}$  & 4p $^2$P$^\mathrm{o}_{1/2}$  & \\
      1165&674 & \Ion{Fe}{6} & 4s $^2$P$^\mathrm{ }_{3/2}$  & 4p $^2$P$^\mathrm{o}_{3/2}$  & \\
      1167&695 & \Ion{Fe}{6} & 4s $^2$G$^\mathrm{ }_{9/2}$  & 4p $^2$F$^\mathrm{o}_{7/2}$  & \\
      1206&041 & \Ion{Fe}{6} & 4s $^4$P$^\mathrm{ }_{1/2}$  & 4p $^4$P$^\mathrm{o}_{3/2}$  & uncertain \\
      1227&883 & \Ion{Fe}{6} & 4s $^4$P$^\mathrm{ }_{5/2}$  & 4p $^4$P$^\mathrm{o}_{3/2}$  & uncertain \\
      1228&605 & \Ion{Fe}{6} & 4s $^2$G$^\mathrm{ }_{9/2}$  & 4p $^2$H$^\mathrm{o}_{11/2}$ & uncertain \\
      1228&681 & \Ion{Fe}{6} & 4s $^2$D$^\mathrm{ }_{3/2}$  & 4p $^2$D$^\mathrm{o}_{5/2}$  & uncertain \\
      1228&962 & \Ion{Fe}{6} & 4s $^4$F$^\mathrm{ }_{3/2}$  & 4p $^4$D$^\mathrm{o}_{3/2}$  & \\
      1228&977 & \Ion{Fe}{6} & 4s $^2$D$^\mathrm{ }_{5/2}$  & 4p $^4$D$^\mathrm{o}_{7/2}$  & \\
      1232&477 & \Ion{Fe}{6} & 4s $^4$F$^\mathrm{ }_{7/2}$  & 4p $^4$D$^\mathrm{o}_{5/2}$  & \\
      1236&973 & \Ion{Fe}{6} & 4s $^4$F$^\mathrm{ }_{5/2}$  & 4p $^4$D$^\mathrm{o}_{3/2}$  & uncertain \\
      1246&835 & \Ion{Fe}{6} & 4s $^4$F$^\mathrm{ }_{5/2}$  & 4p $^2$F$^\mathrm{o}_{5/2}$  & weak \\
      1252&769 & \Ion{Fe}{6} & 4s $^4$P$^\mathrm{ }_{3/2}$  & 4p $^2$D$^\mathrm{o}_{3/2}$  & \\
      1252&789 & \Ion{Fe}{6} & 4s $^2$F$^\mathrm{ }_{5/2}$  & 4p $^2$G$^\mathrm{o}_{7/2}$  & \\
      1253&045 & \Ion{Fe}{6} & 4s $^2$D$^\mathrm{ }_{5/2}$  & 4p $^4$D$^\mathrm{o}_{5/2}$  & \\
      1253&675 & \Ion{Fe}{6} & 4s $^2$G$^\mathrm{ }_{7/2}$  & 4p $^2$H$^\mathrm{o}_{9/2}$  & blend \Ion{S}{2} i.s. \\
      1258&021 & \Ion{Fe}{6} & 4s $^4$F$^\mathrm{ }_{9/2}$  & 4p $^2$F$^\mathrm{o}_{7/2}$  & \\
      1258&881 & \Ion{Fe}{6} & 4s $^4$F$^\mathrm{ }_{7/2}$  & 4p $^2$F$^\mathrm{o}_{5/2}$  & \\
      1260&746 & \Ion{Fe}{6} & 4s $^4$P$^\mathrm{ }_{5/2}$  & 4p $^4$D$^\mathrm{o}_{7/2}$  & blend \Ion{C}{1}, \Ion{Si}{2} i.s. \\
      1261&058 & \Ion{Fe}{6} & 4s $^4$F$^\mathrm{ }_{7/2}$  & 4p $^4$F$^\mathrm{o}_{9/2}$  & \\
      1265&872 & \Ion{Fe}{6} & 4s $^4$F$^\mathrm{ }_{5/2}$  & 4p $^4$F$^\mathrm{o}_{7/2}$  & \\
      1266&103 & \Ion{Fe}{6} & 4s $^2$D$^\mathrm{ }_{5/2}$  & 4p $^2$F$^\mathrm{o}_{7/2}$  & \\
      1271&099 & \Ion{Fe}{6} & 4s $^4$P$^\mathrm{ }_{3/2}$  & 4p $^4$D$^\mathrm{o}_{5/2}$  & \\
      1272&066 & \Ion{Fe}{6} & 4s $^4$F$^\mathrm{ }_{9/2}$  & 4p $^4$G$^\mathrm{o}_{11/2}$ & \\
      1273&843 & \Ion{Fe}{6} & 4s $^4$F$^\mathrm{ }_{3/2}$  & 4p $^4$F$^\mathrm{o}_{5/2}$  & \\
      1276&877 & \Ion{Fe}{6} & 4s $^4$F$^\mathrm{ }_{9/2}$  & 4p $^4$F$^\mathrm{o}_{9/2}$  & \\
      1277&069 & \Ion{Fe}{6} & 4s $^4$P$^\mathrm{ }_{1/2}$  & 4p $^4$D$^\mathrm{o}_{3/2}$  & \\
      1278&292 & \Ion{Fe}{6} & 4s $^4$F$^\mathrm{ }_{7/2}$  & 4p $^4$F$^\mathrm{o}_{7/2}$  & \\
      1282&452 & \Ion{Fe}{6} & 4s $^4$F$^\mathrm{ }_{5/2}$  & 4p $^4$F$^\mathrm{o}_{5/2}$  & \\
      1285&362 & \Ion{Fe}{6} & 4s $^4$F$^\mathrm{ }_{7/2}$  & 4p $^4$G$^\mathrm{o}_{9/2}$  & \\
      1286&088 & \Ion{Fe}{6} & 4s $^4$P$^\mathrm{ }_{5/2}$  & 4p $^4$D$^\mathrm{o}_{5/2}$  & \\
      1286&235 & \Ion{Fe}{6} & 4s $^4$P$^\mathrm{ }_{3/2}$  & 4p $^4$D$^\mathrm{o}_{3/2}$  & \\
      1287&028 & \Ion{Fe}{6} & 4s $^4$F$^\mathrm{ }_{3/2}$  & 4p $^4$F$^\mathrm{o}_{3/2}$  & \\
                        1291&437 & \Ion{Fe}{6} & 4s $^2$D$^\mathrm{ }_{5/2}$  & 4p $^2$F$^\mathrm{o}_{5/2}$  & \\
                        1292&643 & \Ion{Fe}{6} & 4s $^4$P$^\mathrm{ }_{3/2}$  & 4p $^4$D$^\mathrm{o}_{1/2}$  & \\
      1294&549 & \Ion{Fe}{6} & 4s $^4$F$^\mathrm{ }_{9/2}$  & 4p $^4$F$^\mathrm{o}_{7/2}$  & \\
      1295&201 & \Ion{Fe}{6} & 4s $^4$F$^\mathrm{ }_{7/2}$  & 4p $^4$F$^\mathrm{o}_{5/2}$  & \\
      1295&817 & \Ion{Fe}{6} & 4s $^4$F$^\mathrm{ }_{5/2}$  & 4p $^4$F$^\mathrm{o}_{3/2}$  & \\
      1296&734 & \Ion{Fe}{6} & 4s $^2$D$^\mathrm{ }_{3/2}$  & 4p $^2$F$^\mathrm{o}_{5/2}$  & \\
      1296&872 & \Ion{Fe}{6} & 4s $^4$F$^\mathrm{ }_{5/2}$  & 4p $^4$G$^\mathrm{o}_{7/2}$  & \\
                        1299&848 & \Ion{Fe}{6} & 4s $^4$P$^\mathrm{ }_{5/2}$  & 4p $^2$F$^\mathrm{o}_{7/2}$  & \\
      1301&174 & \Ion{Fe}{6} & 4s $^2$D$^\mathrm{ }_{5/2}$  & 4p $^2$P$^\mathrm{o}_{3/2}$  & \\
      1301&800 & \Ion{Fe}{6} & 4s $^4$F$^\mathrm{ }_{9/2}$  & 4p $^4$G$^\mathrm{o}_{9/2}$  & \\
      1308&644 & \Ion{Fe}{6} & 4s $^4$F$^\mathrm{ }_{3/2}$  & 4p $^4$G$^\mathrm{o}_{5/2}$  & \\
      1309&910 & \Ion{Fe}{6} & 4s $^4$F$^\mathrm{ }_{7/2}$  & 4p $^4$G$^\mathrm{o}_{7/2}$  & \\
      1317&731 & \Ion{Fe}{6} & 4s $^4$F$^\mathrm{ }_{5/2}$  & 4p $^4$G$^\mathrm{o}_{5/2}$  & \\
      1324&286 & \Ion{Fe}{6} & 4s $^2$F$^\mathrm{ }_{5/2}$  & 4p $^2$D$^\mathrm{o}_{3/2}$  & \\
      1329&177 & \Ion{Fe}{6} & 4s $^2$F$^\mathrm{ }_{7/2}$  & 4p $^2$D$^\mathrm{o}_{5/2}$  & \\
      1330&971 & \Ion{Fe}{6} & 4s $^2$F$^\mathrm{ }_{5/2}$  & 4p $^4$D$^\mathrm{o}_{5/2}$  & \\
      1336&839 & \Ion{Fe}{6} & 4s $^4$P$^\mathrm{ }_{5/2}$  & 4p $^2$P$^\mathrm{o}_{3/2}$  & \\
      1337&692 & \Ion{Fe}{6} & 4s $^2$D$^\mathrm{ }_{5/2}$  & 4p $^4$S$^\mathrm{o}_{3/2}$  & \\
      1337&793 & \Ion{Fe}{6} & 4s $^2$F$^\mathrm{ }_{7/2}$  & 4p $^4$D$^\mathrm{o}_{7/2}$  & \\
      1361&817 & \Ion{Fe}{6} & 4s $^2$F$^\mathrm{ }_{5/2}$  & 4p $^2$F$^\mathrm{o}_{5/2}$  & \\
                        \noalign{\smallskip}
                        \hline
                \end{tabular}
        \end{center}
        \label{tab:Element_Abundances:Iron2}
\end{table}

\begin{table}
        \caption{Nickel lines identified in the HST/COS spectra of \wdnull and \pgnull.}
\tiny
        \begin{center}
                \begin{tabular}{r@{.}lcccl}
                        \hline
                        \hline
                        \noalign{\smallskip}
                        \multicolumn{2}{c}{Wavelength\,/\,\AA} & \multirow{2}{*}{Ion} & \multicolumn{2}{c}{Transition} & \multirow{2}{*}{Comment} \\
                        \cline{1-2}
                        \cline{4-5}
                        \noalign{\smallskip}
                        \multicolumn{2}{c}{laboratory}         &                            & low & up                       & \\
                        \noalign{\smallskip}
                        \hline
                        \noalign{\smallskip}
                        1357&064                    & \Ion{Ni}{4} & 4s $^6$D$^\mathrm{ }_{9/5}$  & 4p $^6$P$^\mathrm{o}_{7/2}$  & weak \\
                        \two                        &             &                              &                              & \\
      1231&875                    & \Ion{Ni}{5} & 4s $^3$D$^\mathrm{ }_{3}$    & 4p $^3$P$^\mathrm{o}_{2}$    & weak \\
                        1232&524                    & \Ion{Ni}{5} & 4s $^5$G$^\mathrm{ }_{4}$    & 4p $^5$F$^\mathrm{o}_{3}$    & weak \\
                        1232&807                    & \Ion{Ni}{5} & 4s $^1$H$^\mathrm{ }_{5}$    & 4p $^1$H$^\mathrm{o}_{5}$    & weak \\
                        1232&964                    & \Ion{Ni}{5} & 4s $^3$D$^\mathrm{ }_{3}$    & 4p $^3$F$^\mathrm{o}_{4}$    & weak \\
                        1233&257                    & \Ion{Ni}{5} & 4s $^5$F$^\mathrm{ }_{5}$    & 4p $^5$F$^\mathrm{o}_{5}$    & weak \\
                        1233&312                    & \Ion{Ni}{5} & 4s $^5$G$^\mathrm{ }_{3}$    & 4p $^5$F$^\mathrm{o}_{2}$    & weak \\
                        1235&831                    & \Ion{Ni}{5} & 4s $^1$I$^\mathrm{ }_{6}$    & 4p $^5$G$^\mathrm{o}_{6}$    & weak \\
                        1242&043                    & \Ion{Ni}{5} & 4s $^1$D$^\mathrm{ }_{2}$    & 4p $^3$G$^\mathrm{o}_{3}$    & \\
                        1244&174\,\tablefootmark{a} & \Ion{Ni}{5} & 4s $^7$S$^\mathrm{ }_{3}$    & 4p $^7$P$^\mathrm{o}_{4}$    & \\
                        1245&020                    & \Ion{Ni}{5} & 4s $^1$F$^\mathrm{ }_{3}$    & 4p $^3$F$^\mathrm{o}_{3}$    & weak \\
                        1245&074                    & \Ion{Ni}{5} & 4s $^3$G$^\mathrm{ }_{4}$    & 4p $^3$F$^\mathrm{o}_{4}$    & weak \\
                        1245&203                    & \Ion{Ni}{5} & 4s $^3$D$^\mathrm{ }_{2}$    & 4p $^3$F$^\mathrm{o}_{3}$    & weak\\
                        1249&522                    & \Ion{Ni}{5} & 4s $^5$G$^\mathrm{ }_{5}$    & 4p $^5$F$^\mathrm{o}_{4}$    & \\
                        1250&033                    & \Ion{Ni}{5} & 4s $^5$G$^\mathrm{ }_{4}$    & 4p $^5$F$^\mathrm{o}_{4}$    & weak \\
                        1251&812\,\tablefootmark{a} & \Ion{Ni}{5} & 4s $^3$G$^\mathrm{ }_{5}$    & 4p $^3$G$^\mathrm{o}_{5}$    & \\
                        1252&183\,\tablefootmark{a} & \Ion{Ni}{5} & 4s $^5$G$^\mathrm{ }_{6}$    & 4p $^5$F$^\mathrm{o}_{5}$    & \\
                        1253&980                    & \Ion{Ni}{5} & 4s $^3$G$^\mathrm{ }_{3}$    & 4p $^3$G$^\mathrm{o}_{3}$    & \\
                        1257&626\,\tablefootmark{a} & \Ion{Ni}{5} & 4s $^5$G$^\mathrm{ }_{5}$    & 4p $^5$H$^\mathrm{o}_{6}$    & \\
                        1261&327                    & \Ion{Ni}{5} & 4s $^3$F$^\mathrm{ }_{4}$    & 4p $^3$G$^\mathrm{o}_{5}$    & \\
                        1261&760                    & \Ion{Ni}{5} & 4s $^5$D$^\mathrm{ }_{4}$    & 4p $^5$F$^\mathrm{o}_{5}$    & \\
                        1264&501\,\tablefootmark{a} & \Ion{Ni}{5} & 4s $^7$S$^\mathrm{ }_{3}$    & 4p $^7$P$^\mathrm{o}_{3}$    & \\
                        1265&671\,\tablefootmark{b} & \Ion{Ni}{5} & 4s $^3$H$^\mathrm{ }_{6}$    & 4p $^3$I$^\mathrm{o}_{7}$    & blend with unid. \\
                        1265&725\,\tablefootmark{b} & \Ion{Ni}{5} & 4s $^5$F$^\mathrm{ }_{3}$    & 4p $^5$G$^\mathrm{o}_{4}$    & blend with unid. \\                  
                        1266&408\,\tablefootmark{a} & \Ion{Ni}{5} & 4s $^5$G$^\mathrm{ }_{4}$    & 4p $^5$H$^\mathrm{o}_{5}$    & \\
                        1270&677                    & \Ion{Ni}{5} & 4s $^3$I$^\mathrm{ }_{7}$    & 4p $^3$K$^\mathrm{o}_{8}$    & \\
                        1273&204                    & \Ion{Ni}{5} & 4s $^5$G$^\mathrm{ }_{3}$    & 4p $^5$H$^\mathrm{o}_{4}$    & \\
                        1273&827                    & \Ion{Ni}{5} & 4s $^5$P$^\mathrm{ }_{2}$    & 4p $^5$P$^\mathrm{o}_{3}$    & weak \\
                        1276&428\,\tablefootmark{a} & \Ion{Ni}{5} & 4s $^5$D$^\mathrm{ }_{3}$    & 4p $^5$F$^\mathrm{o}_{4}$    & \\
                        1276&958\,\tablefootmark{a} & \Ion{Ni}{5} & 4s $^7$S$^\mathrm{ }_{3}$    & 4p $^7$P$^\mathrm{o}_{2}$    & \\
                        1279&325                    & \Ion{Ni}{5} & 4s $^5$P$^\mathrm{ }_{3}$    & 4p $^5$D$^\mathrm{o}_{4}$    & weak \\
                        1279&720                    & \Ion{Ni}{5} & 4s $^5$G$^\mathrm{ }_{2}$    & 4p $^5$H$^\mathrm{o}_{3}$    & \\
                        1280&138                    & \Ion{Ni}{5} & 4s $^3$H$^\mathrm{ }_{4}$    & 4p $^3$I$^\mathrm{o}_{5}$    & weak \\
                        1282&201                    & \Ion{Ni}{5} & 4s $^3$I$^\mathrm{ }_{6}$    & 4p $^3$I$^\mathrm{o}_{6}$    & weak \\
                        1282&270                    & \Ion{Ni}{5} & 4s $^3$I$^\mathrm{ }_{5}$    & 4p $^3$I$^\mathrm{o}_{6}$    & weak \\
                        1282&724                    & \Ion{Ni}{5} & 4s $^3$I$^\mathrm{ }_{7}$    & 4p $^3$I$^\mathrm{o}_{6}$    & weak \\
                        1287&553                    & \Ion{Ni}{5} & 4s $^3$D$^\mathrm{ }_{3}$    & 4p $^3$F$^\mathrm{o}_{4}$    & weak \\
                        1287&628                    & \Ion{Ni}{5} & 4s $^1$D$^\mathrm{ }_{2}$    & 4p $^1$F$^\mathrm{o}_{3}$    & weak \\
                        1287&808                    & \Ion{Ni}{5} & 4s $^5$D$^\mathrm{ }_{1}$    & 4p $^5$F$^\mathrm{o}_{2}$    & weak \\
                        1300&979                    & \Ion{Ni}{5} & 4s $^5$S$^\mathrm{ }_{2}$    & 4p $^5$P$^\mathrm{o}_{1}$    & \\
                        1304&870                    & \Ion{Ni}{5} & 4s $^3$I$^\mathrm{ }_{5}$    & 4p $^3$I$^\mathrm{o}_{5}$    & \\
                        1305&696                    & \Ion{Ni}{5} & 4s $^3$I$^\mathrm{ }_{6}$    & 4p $^3$K$^\mathrm{o}_{7}$    & weak \\
                        1306&238                    & \Ion{Ni}{5} & 4s $^3$I$^\mathrm{ }_{7}$    & 4p $^3$K$^\mathrm{o}_{7}$    & weak \\
                        1306&624                    & \Ion{Ni}{5} & 4s $^5$G$^\mathrm{ }_{6}$    & 4p $^5$G$^\mathrm{o}_{6}$    & \\
                        1307&603                    & \Ion{Ni}{5} & 4s $^5$S$^\mathrm{ }_{2}$    & 4p $^5$P$^\mathrm{o}_{2}$    & \\
                        1311&106                    & \Ion{Ni}{5} & 4s $^5$G$^\mathrm{ }_{5}$    & 4p $^5$G$^\mathrm{o}_{5}$    & \\
                        1312&718\,\tablefootmark{a} & \Ion{Ni}{5} & 4s $^3$I$^\mathrm{ }_{5}$    & 4p $^3$K$^\mathrm{o}_{6}$    & \\
                        1313&280                    & \Ion{Ni}{5} & 4s $^5$G$^\mathrm{ }_{4}$    & 4p $^5$G$^\mathrm{o}_{4}$    & \\
                        1314&330\,\tablefootmark{a} & \Ion{Ni}{5} & 4s $^5$G$^\mathrm{ }_{3}$    & 4p $^5$G$^\mathrm{o}_{3}$    & \\
                        1314&349                    & \Ion{Ni}{5} & 4s $^1$H$^\mathrm{ }_{5}$    & 4p $^1$I$^\mathrm{o}_{6}$    & \\
                        1314&682                    & \Ion{Ni}{5} & 4s $^5$G$^\mathrm{ }_{2}$    & 4p $^5$G$^\mathrm{o}_{2}$    & weak \\
                        1317&447                    & \Ion{Ni}{5} & 4s $^1$I$^\mathrm{ }_{6}$    & 4p $^1$K$^\mathrm{o}_{7}$    & \\
                        1318&327                    & \Ion{Ni}{5} & 4s $^3$D$^\mathrm{ }_{2}$    & 4p $^3$F$^\mathrm{o}_{3}$    & weak \\
                        1318&515\,\tablefootmark{a} & \Ion{Ni}{5} & 4s $^5$S$^\mathrm{ }_{2}$    & 4p $^5$P$^\mathrm{o}_{3}$    & \\
                        1323&562                    & \Ion{Ni}{5} & 4s $^3$H$^\mathrm{ }_{6}$    & 4p $^3$H$^\mathrm{o}_{6}$    & weak \\
                        1323&977                    & \Ion{Ni}{5} & 4s $^3$G$^\mathrm{ }_{3}$    & 4p $^3$H$^\mathrm{o}_{4}$    & \\
                        1329&358\,\tablefootmark{a} & \Ion{Ni}{5} & 4s $^3$G$^\mathrm{ }_{4}$    & 4p $^3$H$^\mathrm{o}_{5}$    & \\
                        1336&136\,\tablefootmark{a} & \Ion{Ni}{5} & 4s $^3$G$^\mathrm{ }_{5}$    & 4p $^3$H$^\mathrm{o}_{6}$    & \\
                        1347&720                    & \Ion{Ni}{5} & 4s $^3$G$^\mathrm{ }_{4}$    & 4p $^3$F$^\mathrm{o}_{3}$    & weak \\
                        \noalign{\smallskip}
                        \hline
                \end{tabular}
                \tablefoot{
                \tablefoottext{a}{also seen in PG\,0109$+$111},
                \tablefoottext{b}{only seen in PG\,0109$+$111}
                }
        \end{center}
        \label{tab:nickel}
\end{table}

\begin{figure}
\centering
\resizebox{0.75\columnwidth}{!}{\includegraphics{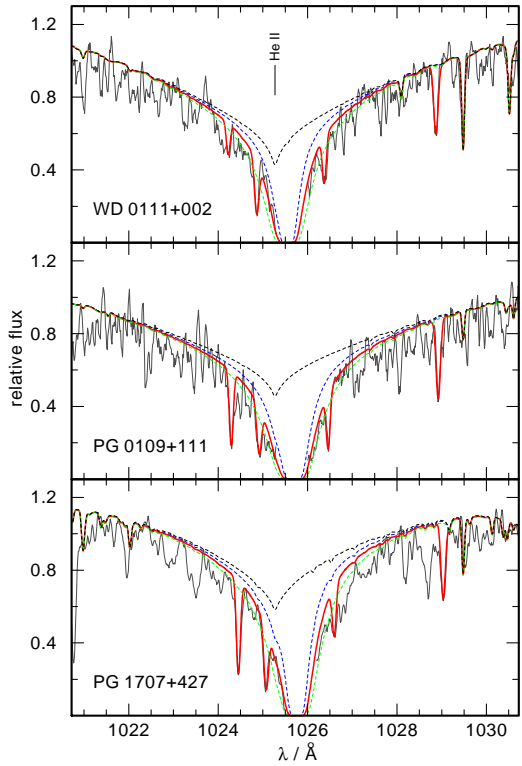}}
\caption{Synthetic spectra around $\mathrm{Ly}\,\beta$ from our final
  models (black, dashed graphs) for the program stars compared with the FUSE observations
  (black). Red graphs: Interstellar
  absorption with neutral hydrogen column densities as given in
  Sect.\,\ref{sect:H_Column_Density} applied. The adjacent dashed
  graphs indicate the error limits.}
\label{fig:H_Column_Density:Hbeta}
\end{figure}

\begin{figure}
\centering
\resizebox{0.75\columnwidth}{!}{\includegraphics{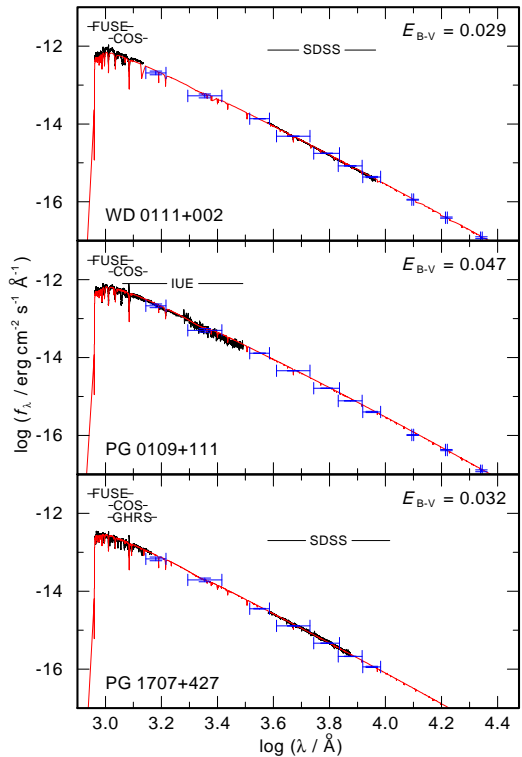}}
\caption{Spectral energy distribution of our program stars compared to
  the final, reddened models (red graphs). Models of \wdnull, and
  \pgnull were normalized to the 2MASS H magnitude and that of
  \pgsieben to the SDSS z magnitude. Magnitudes from GALEX, SDSS, and
  2MASS are plotted in blue with their error bars.}
\label{fig:H_Column_Density:EBV}
\end{figure}

\begin{figure*}
\centering
\resizebox{0.68\textwidth}{!}{\includegraphics{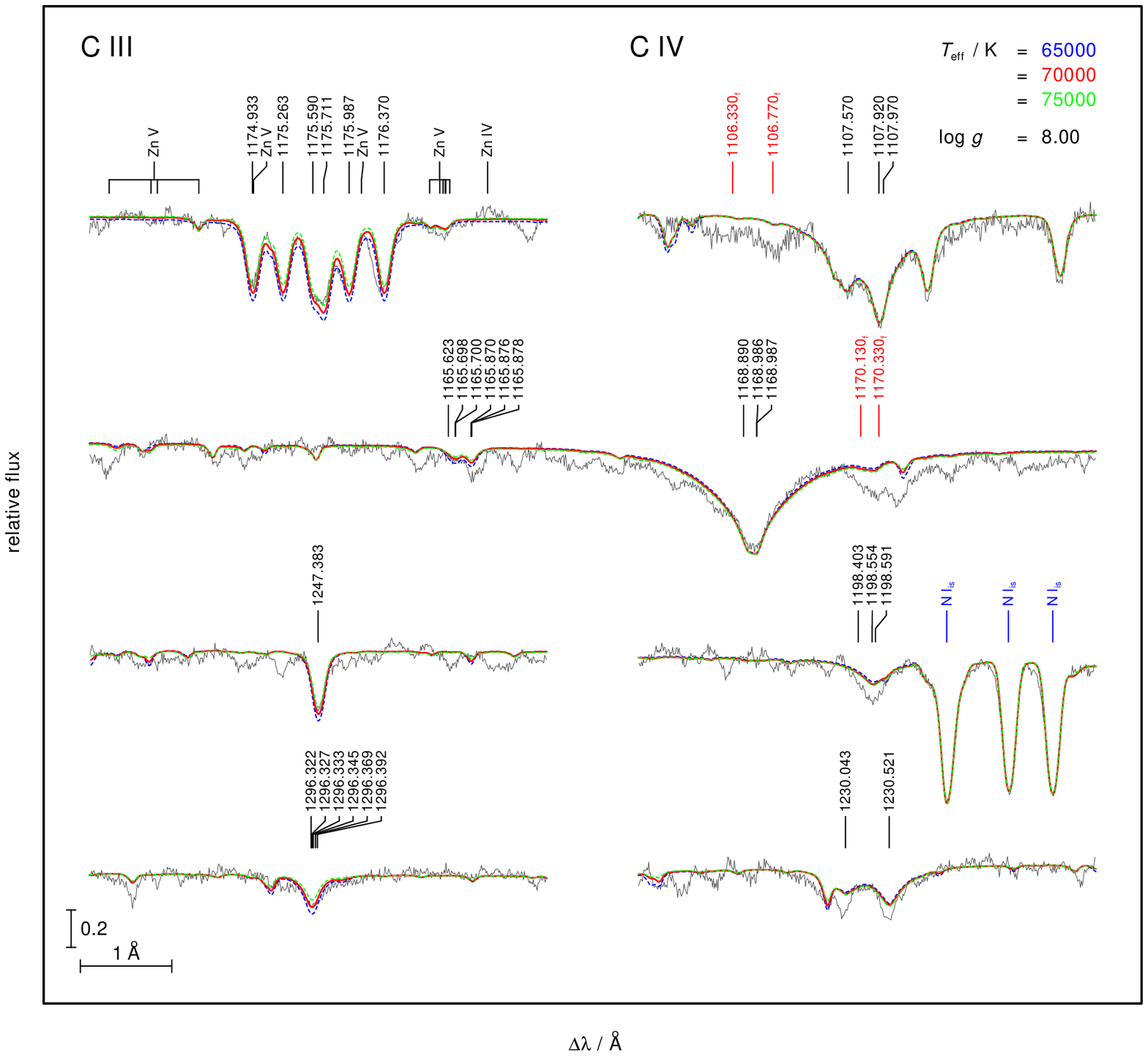}}
\caption{Like Fig.\,\ref{fig:WD0111+002_Teff}, but for \pgnull.}
\label{fig:PG0109+111_Teff}
\end{figure*}

\begin{figure*}
\centering
\resizebox{0.68\textwidth}{!}{\includegraphics{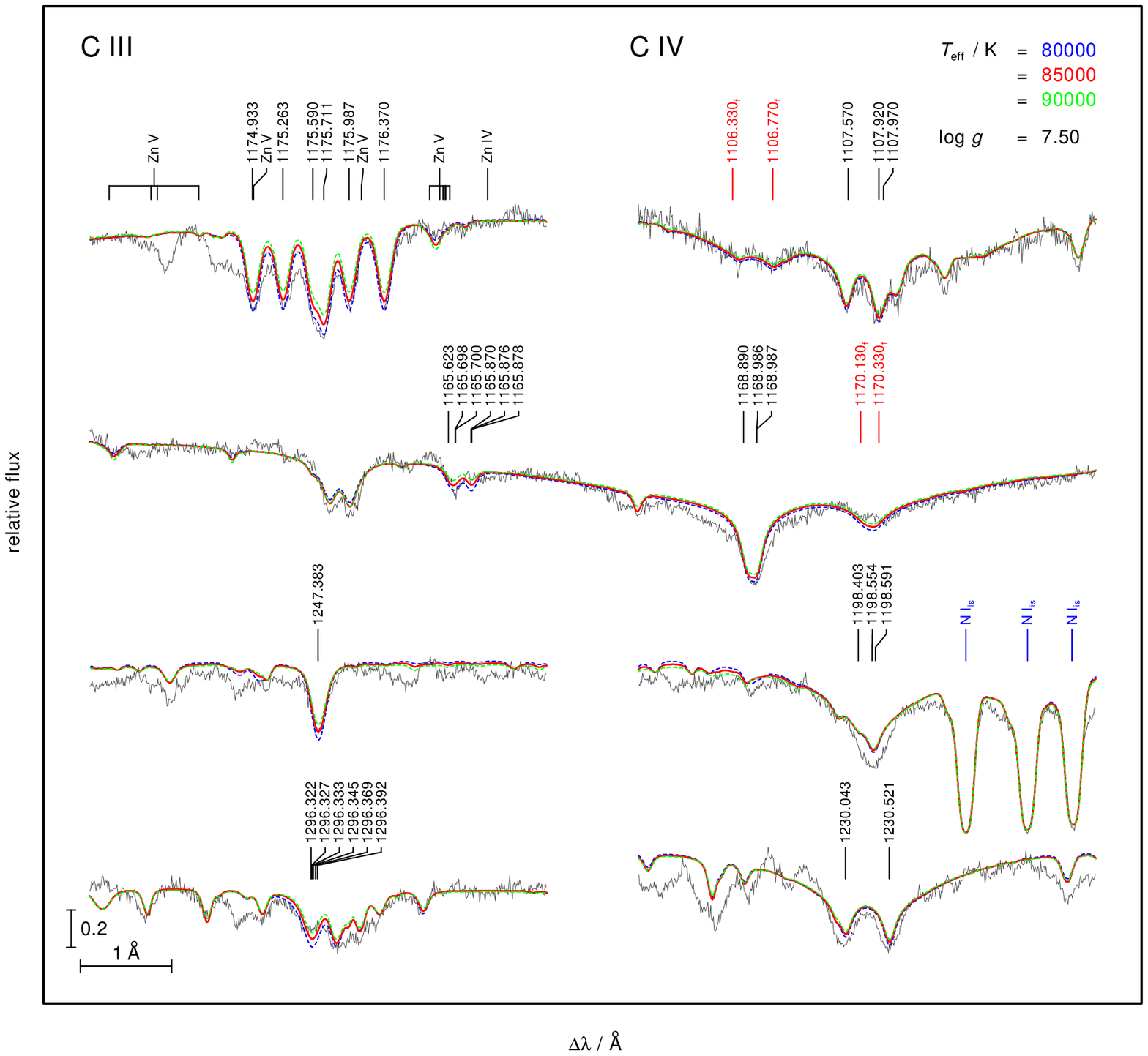}}
\caption{Like Fig.\,\ref{fig:WD0111+002_Teff}, but for \pgsieben.}
\label{fig:PG1707+427_Teff}
\end{figure*}

\begin{figure*}
\centering
\resizebox{0.88\textwidth}{!}{\includegraphics{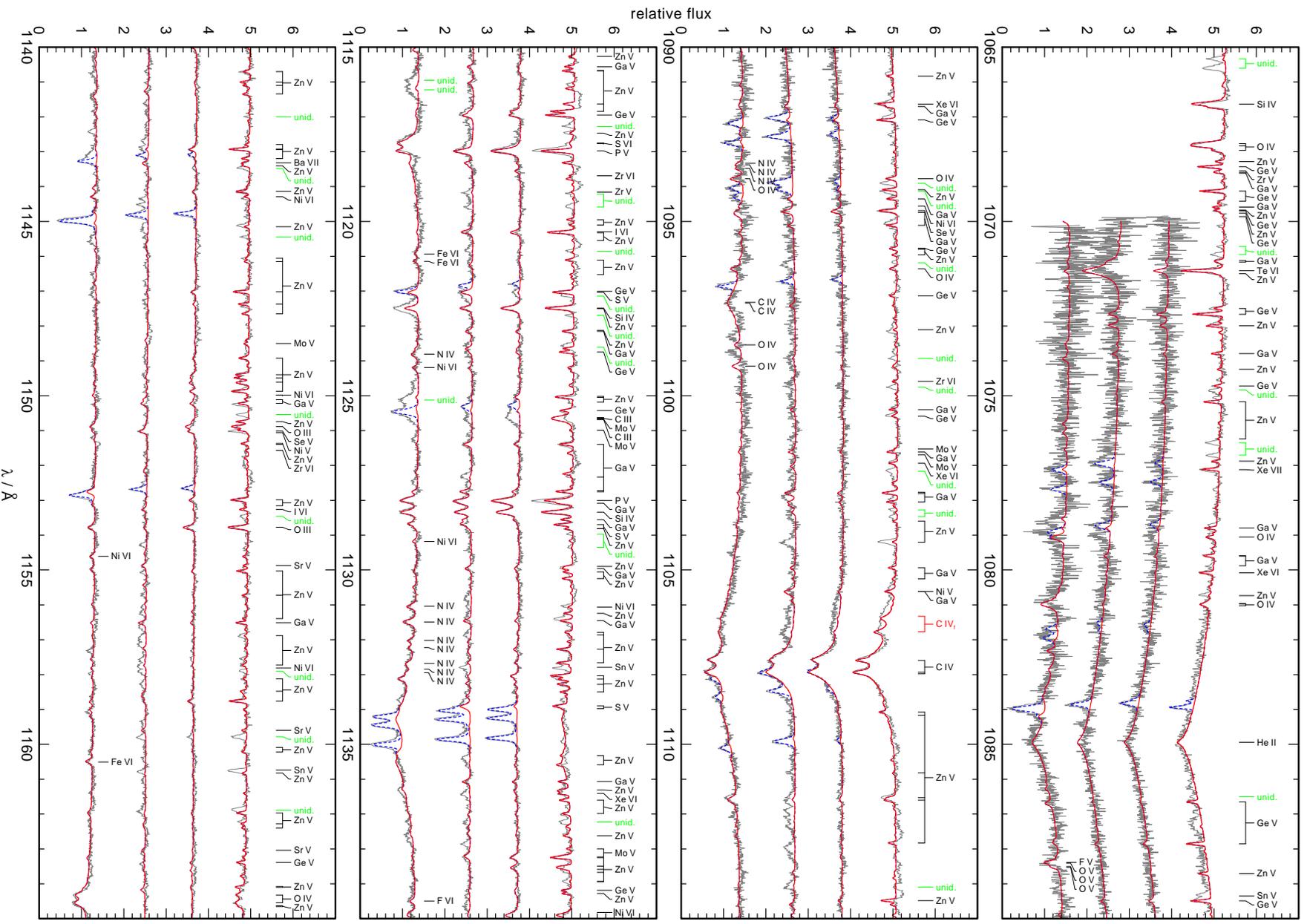}}
\caption{Complete HST/COS spectra of our three program stars (three
  bottom spectra in each panel) together with the HST/STIS spectrum of the
  comparison DO star \re (top spectrum). Overplotted (red graphs) are
  our final models and the final model for \re from
  \citet{rauch17b}. Blue dashed graphs indicate ISM lines.}
\label{fig:all_spectra}
\end{figure*}

\addtocounter{figure}{-1}
\begin{figure*}
\centering
\resizebox{0.88\textwidth}{!}{\includegraphics{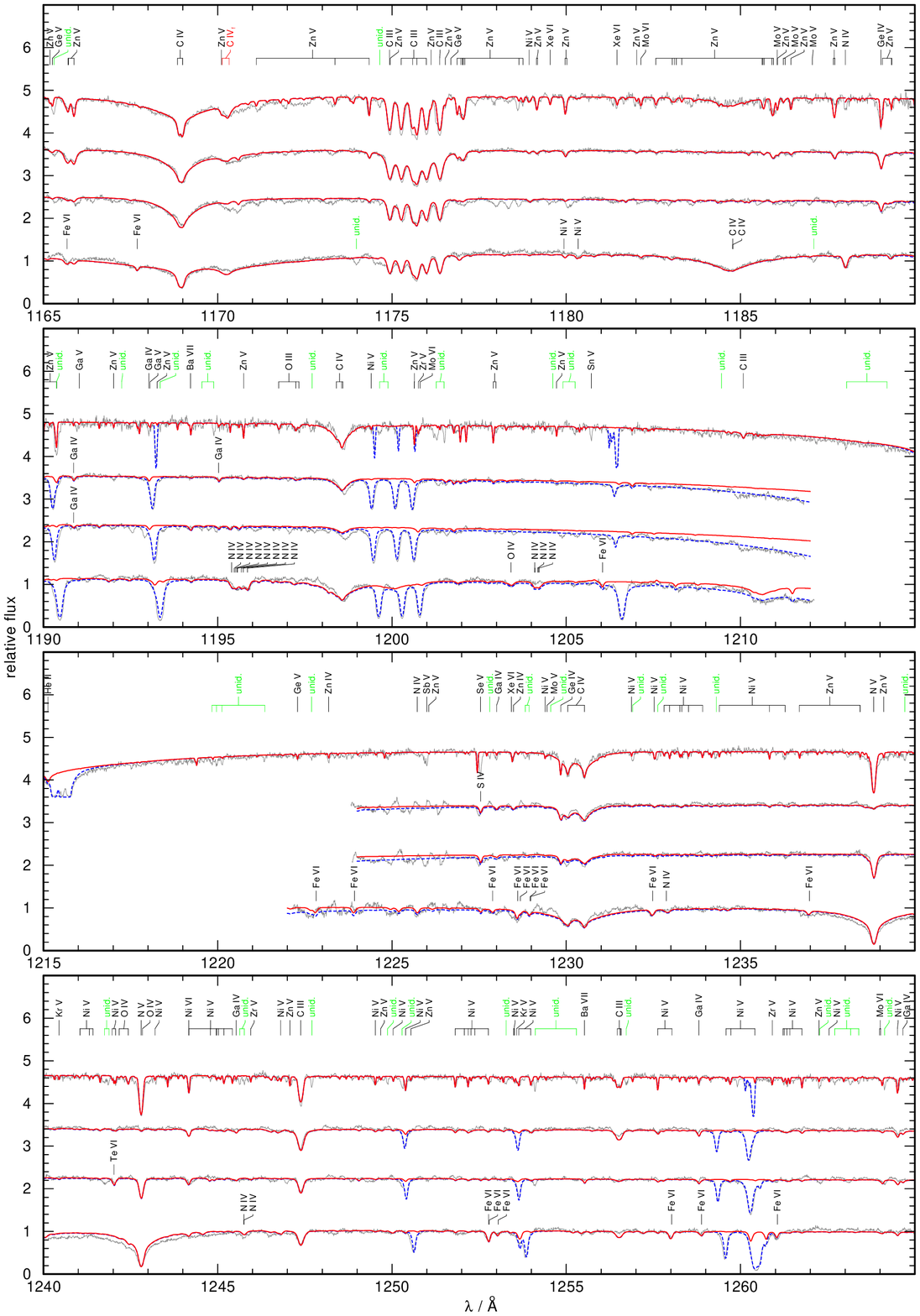}}
\caption{continued. }
\end{figure*}

\addtocounter{figure}{-1}
\begin{figure*}
\centering
\resizebox{0.88\textwidth}{!}{\includegraphics{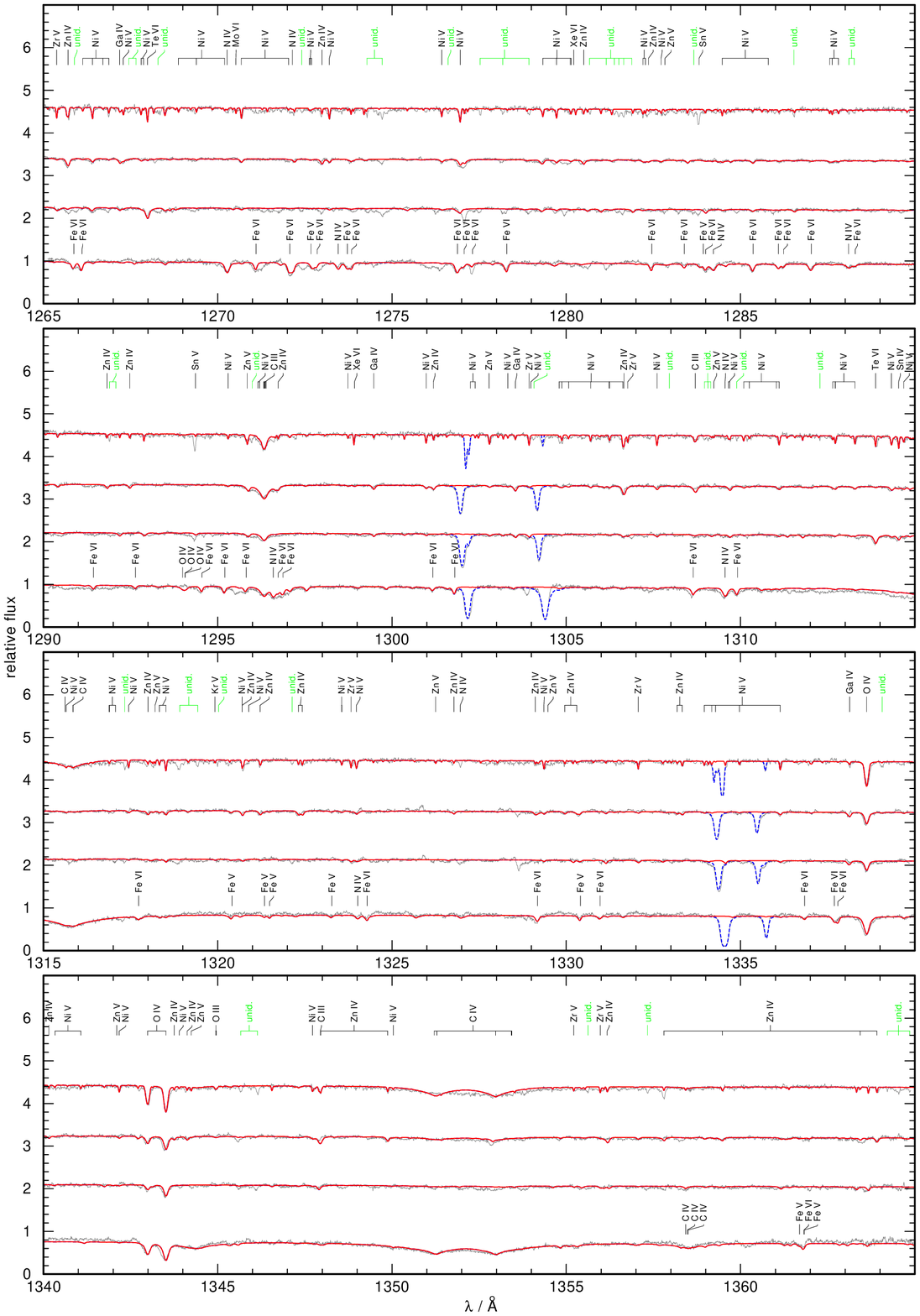}}
\caption{continued. }
\end{figure*}

\end{document}